\documentclass[preprint,12pt]{elsarticle}

\usepackage{booktabs}
\usepackage{amsmath,amssymb,amsthm}
\usepackage{hyperref}
\usepackage{array}
\usepackage{multirow}
\usepackage{url}
\usepackage{lineno}
\usepackage{tikz}
\usetikzlibrary{positioning,arrows.meta,shapes.geometric}

\journal{Expert Systems with Applications}

\begin{document}

\begin{frontmatter}

\title{TourMart: A Parametric Audit Instrument for Commission Steering in LLM Travel Agents}

\author[affA,affB]{Yao Liu\corref{cor1}}
\ead{liuyao@student.usm.my}
\ead[url]{https://orcid.org/0009-0009-3128-7802}

\cortext[cor1]{Corresponding author. ORCID:
\href{https://orcid.org/0009-0009-3128-7802}{0009-0009-3128-7802}.}

\affiliation[affA]{organization={Department of Management and Media, The Engineering and Technology College, Chengdu University of Technology},
              city={Leshan},
              postcode={614000},
              country={China}}

\affiliation[affB]{organization={School of Computer Sciences, Universiti Sains Malaysia},
              city={Penang},
              postcode={11800},
              country={Malaysia}}

\begin{abstract}
Online travel agents---Booking, Trip.com, Expedia---have replaced their
ranked-list interfaces with conversational LLM agents that compress a
page of options into one sentence of prose advice. Each booking earns
the OTA commission, and different hotels and airlines pay different
rates: the agent has a structural incentive to favor higher-margin
recommendations. Whether any specific deployed agent does this, and by
how much, no one can currently measure. Existing tools---disclosure
banners, conversion A/B testing, UI dark-pattern taxonomies, generic
LLM safety scores---were built for older interfaces and miss the
prose-recommendation surface where the steering happens.

We propose \textbf{TourMart}, an applied intelligent-system audit
instrument for LLM-OTA commission governance. Two interpretable
governance levers---$\lambda$ (the weight of message-induced perception
in the traveler's accept/reject decision) and $\kappa$ (a
budget-normalized cap on how far the message can shift perceived
welfare)---drive a paired counterfactual procedure: holding the same
traveler and the same bundle fixed, the steering delta is read off
between a commission-aware OTA prompt and a minimum-disclosure factual
template. A symmetric six-gate producer audit separates LLM-engineering
failures (template collapse, refusal, internal-ID leakage) from genuine
commercial steering.

At deployed governance settings $(\lambda{=}1, \kappa{=}0.05)$, a
Qwen-14B reader shows $+7.69$pp commission-induced steering (exact
McNemar $p{=}0.003$); a Llama-3.1-8B reader shows $+3.50$pp in the same
direction at $n{=}143$, with a supplemental extended-$n$ replay
($n{=}270$, diagnostic window) confirming significance
($+2.96$pp, $p{=}0.008$). Across the $(\lambda, \kappa)$ governance grid
both readers pass family-wise scenario-clustered statistical correction
(Qwen $p{<}0.001$, Llama $p{=}0.008$). TourMart's output is a sentence a
compliance report can quote: \emph{``at this deployment, 7.7 extra
commission-steered recommendations per 100 paired traveler sessions.''}
\end{abstract}


\end{frontmatter}


\section{Introduction}
\label{sec:intro}

A VP of Compliance at an online travel aggregator is preparing a board
update six months after her company shipped an LLM-based travel assistant.
The dashboard she can show the board is conversion: up 3.1\% over the
prior quarter. The disclosure she can point to is a two-line banner at the
bottom of the chat window. The audit committee asks the obvious question:
does our assistant steer travelers toward higher-commission inventory? She
cannot answer. This scene is not hypothetical---Ctrip's ``big-data price
discrimination'' controversies in the early 2020s placed OTA recommendation
governance on Chinese regulators' files, and tourism bookings run roughly
RMB 5{,}000--15{,}000 for domestic and RMB 20{,}000--50{,}000+ for
international packages, are purchased at low frequency, and are
information-asymmetric in ways that e-commerce of everyday goods is not.
A mis-steered booking is not meaningfully returnable. \textit{For a given
LLM travel-agent deployment at a given configuration, how much does it
steer? That is the question this paper answers.} TourMart, the applied
\textbf{intelligent-system audit instrument} we propose, takes a deployed
LLM-OTA configuration as input and outputs a single
deployment-specific number the audit committee can quote.

Four tools already sit on the compliance officer's shelf, and none of
them gives her the answer.

\textit{Disclosure banners} (``responses generated with AI assistance'')
were designed for ranked listings, where a user can scan ten options and
discount whichever is tagged sponsored. When the LLM compresses those ten
options into one sentence, that tag has nowhere to
attach~\cite{ftc2023endorsement, eu2022dsa}.

\textit{Conversion A/B testing} tells the platform whether a configuration
sells, not whether it is honest. A configuration that confidently steers
a traveler toward a higher-margin mismatch produces the same conversion
lift as one that surfaces a genuinely better
match~\cite{aridor2022recommenderoriginals}.

\textit{Dark-pattern taxonomies}~\cite{gray2018darkpatterns,
gray2024ontology, mathur2019darkpatterns} catalog UI manipulation
techniques (scarcity timers, forced-confirmation dialogs, deceptive
button placement) built for the era when the manipulation lived in
buttons. The LLM-OTA setting moves the manipulation into a single
sentence of prose, where these taxonomies do not reach.

\textit{Generic LLM safety scoring}---helpfulness, harmlessness,
refusal-to-comply---audits whether the model would tell a user how to
make a bomb. It has no commission-awareness, no welfare anchor, and no
comparison to a counterfactual message; it cannot tell the audit
committee whether the model favors suppliers who pay the platform more.

Regulatory frameworks across jurisdictions---EU DSA
Article~25~\cite{eu2022dsa}, FTC 16~CFR~255~\cite{ftc2023endorsement},
MiFID~II Article~27~\cite{eu2014mifid2}, EU P2B Regulation
2019/1150~\cite{eu2019p2b}, and China's 2022 Algorithm Recommendation
Provisions (CAC Order No.~9)~\cite{china2022algorec}---each name the harm
(``manipulation,'' ``deception,'' ``undue interference with consumer
choice'') and prohibit it in algorithmically-mediated commercial
contexts. None of them tells the compliance officer how to measure
whether her own LLM is doing it. \textit{There is no fuel-economy label
for LLM-OTA deployments.}

TourMart supplies the missing instrument through a three-step audit
design. \textbf{(1) Paired counterfactual replay.} The same traveler
agent evaluates the same bundles twice: once under a commission-aware
OTA message (the commercial context, where the LLM knows it sells for
margin), and once under a minimum-disclosure factual template (no sales
language, just the bundle's objective specifications). The acceptance
gap between these two runs, with bundle and traveler held fixed, is the
steering reading. \textbf{(2) Two governance dials swept across a
$\boldsymbol{6{\times}6}$ grid.} A gain $\lambda$ and a saturation cap
$\kappa$ act on the message-induced perception term inside a frozen
welfare rule (formal definitions in
\S\ref{sec:formulation}), mapping how steering changes as platform
governance parameters drift away from the deployed point. \textbf{(3) A
symmetric six-gate audit on the producer LLM} that wrote the messages,
separating generator-side failure modes (template collapse, refusal
hedging, internal-ID leakage) from genuine commercial steering.
TourMart's output is not a leaderboard score; it is a sentence the audit
committee can quote: \textit{``at $(\lambda{=}1, \kappa{=}0.05)$, this
deployment produces $+X$pp paired steering under the hardened welfare
rule.''}

We instantiate TourMart with a Qwen-7B-Instruct OTA producer and two
traveler-reader backbones (Qwen-14B-AWQ and Llama-3.1-8B) across 143
near-threshold paired stimuli---the regime where the welfare rule is
mechanically flippable. At the deployed governance point
$(\lambda{=}1, \kappa{=}0.05)$, the Qwen reader shows
\textbf{$+7.69$pp} commission-induced steering (exact McNemar
$p{=}0.003$). The Llama reader shows $+3.50$pp in the same direction; at
$n{=}143$ this arm is under-powered for the pre-registered exact test,
but a supplemental extended-$n$ replay ($n{=}270$, diagnostic window)
confirms significance at $\alpha{=}0.05$
($+2.96$pp, $b/c{=}8/0$, exact $p{=}0.008$; \S\ref{sec:results}.1).
Across the 2D governance grid, peak
steering reaches \textbf{$+10.49$pp} on Qwen and \textbf{$+7.69$pp} on
Llama, with scenario-clustered grid max-stat permutation $p{<}0.001$ on
Qwen and $p{=}0.008$ on Llama. A symmetric six-gate audit on the
producer LLM surfaces two distinct default-prompt failure modes: Qwen
over-hedges ($55.9\%$ refusal under a hardened refusal classifier),
Llama template-collapses with $80.9$--$84.6\%$ internal-ID leakage in
its messages. \textit{These are governance readings, not capability
claims: TourMart audits a deployed configuration, it does not score the
underlying language model.}

\subsection*{Contributions}

\begin{enumerate}
  \item \textbf{A parametric audit instrument for commission governance.}
        Two interpretable governance dials, $\lambda$ and $\kappa$, act on
        a frozen welfare-rule decision to produce per-configuration
        behavioral readings rather than a leaderboard score.
  \item \textbf{Scenario-clustered paired inference with channel
        attribution.} Grid max-stat permutation correction with scenario
        as the cluster unit, plus coefficient-zero attribution that
        localizes observed steering to specific perception channels under
        the frozen rule.
  \item \textbf{Symmetric six-gate producer audit.} Identical gates
        applied to the OTA that produced the messages, diagnosing
        generator-side failures (Qwen over-hedging, Llama
        template-collapse with a repairable style fix) that would
        otherwise confound the behavioral reading.
\end{enumerate}

\section{Related Work}
\label{sec:related}

TourMart sits at the intersection of four research lineages whose
top-venue history establishes the legitimacy of its constituent moves;
we are explicit that two of the four are load-bearing (their methods
constrain ours) and two are paradigm-level support (their existence
licenses the move, but we do not inherit their architectures).

\textbf{(L1, paradigm support) LLM-as-behavioral-subject.} Park et al.'s
Generative Agents (UIST~2023)~\cite{park2023generative}, Horton's Homo
Silicus (NBER~2023)~\cite{horton2023homo}, and Argyle et al.'s
silicon-sample framework (Political
Analysis~2023)~\cite{argyle2023outofone} establish that LLMs can be
treated as behavioral subjects in social-scientific simulation. We
inherit the legitimacy of LLM agents as traveler-reader and OTA-producer
stand-ins, not their memory, planning, or fidelity-validation
architectures.

\textbf{(L2, load-bearing) LLM persuasion measurement.} Salvi et al.\
(Nature Human Behaviour~2025)~\cite{salvi2025persuasiveness} and the
concurrent commercial-steering line~\cite{salvi2026commercial,%
bansal2025magentic} supply the paired-counterfactual measurement frame
that TourMart uses to identify message-induced steering. We extend it
from a binary disclosure manipulation to a continuous
$(\lambda, \kappa)$ governance grid acting on a frozen welfare rule.

\textbf{(L3, load-bearing) Algorithmic platform audit.} Sandvig et al.\
(ICA~2014)~\cite{sandvig2014auditing}, Hannak et al.\
(WWW~2013)~\cite{hannak2013measuring}, and Metaxa et al.\ (Foundations
\& Trends in HCI~2021)~\cite{metaxa2021auditing} define the
external-querying audit instrument: probe an opaque platform with
controlled inputs, observe outputs, infer behavior. TourMart
instantiates this paradigm at the LLM-prose-recommendation surface;
$(\lambda, \kappa)$ are audit knobs in this tradition.

\textbf{(L4, paradigm support) Dark patterns in OTA and LLM contexts.}
Kim et al.\ (Annals of Tourism Research~2021)~\cite{kim2021ota} document
UI-level dark patterns on OTA websites; DarkBench (ICLR~2025
Oral)~\cite{kran2025darkbench} extends the dark-pattern taxonomy to LLM
outputs. TourMart's symmetric six-gate producer audit isolates
generator-side failures (template collapse, refusal, internal-ID
leakage) from genuine commercial steering---a layer adjacent to, not
derived from, these taxonomies.

We do not claim a new lineage; each is well established. Our
contribution is their first instrumented collision in the OTA
commission-steering vertical, where no deployment-specific comparative
audit readout currently exists. We make this scope deliberately narrow:
$(\lambda, \kappa)$ are decision-time governance parameters acting on a
frozen welfare rule, not multi-sided exposure-fairness tradeoffs in the
sense of Wu et al.\ (SIGIR~2022)~\cite{wu2022jmefairness}, who jointly
model consumer- and producer-side exposure for ranking. The instrument
is a comparative audit readout, not a calibrated metrology device, and
we treat traveler-reader behavioral fidelity to real bookers as an open
validation question (\S\ref{sec:limitations}).

\medskip

We organize the rest of this section by the governance question: what
platforms do today to self-regulate LLM-OTA steering; what regulators
have tried to require; what researchers have built as adjacent tools; and
where each falls short of the readout a compliance officer needs.

\subsection{What platforms do today}

Industry practice on LLM-mediated recommendation governance rests on three
legs, none of which yields a per-deployment welfare reading. First,
boilerplate disclosure: a short banner appended to every assistant message.
A product team's job is to minimize the banner's salience; it persists but
the prose it frames steers anyway. Second, internal prompt-engineering
policy---trust-and-safety teams write rules such as ``do not mention
higher-commission inventory first''---but these rules are not externally
auditable, not version-controlled against live deployment, and fail
silently when the model drifts. Third, conversion A/B testing, the
workhorse of platform measurement: A/B captures platform benefit (did the
traveler book?) but cannot separate conversion lift from welfare steering.
A message that steers a user toward a higher-commission mismatch produces
the same A/B signal as one that surfaces a better
match~\cite{aridor2022recommenderoriginals}. Large-scale industrial
deployments report task and conversion metrics under internal
policy; none publishes a welfare-relative instrument.
\textit{No per-deployment welfare-relative readout exists in the public
record at the prose-recommendation level.}

\subsection{What regulators have tried}

Regulatory text names the harm without defining it operationally. The EU
Digital Services Act (Article~25)~\cite{eu2022dsa} prohibits interface
design that ``materially distorts or impairs the ability of recipients to
make free and informed decisions,'' but does not specify a test. The FTC
2022 dark-pattern report~\cite{ftc2022darkpatterns} and 16 CFR Part~255
Endorsement Guides~\cite{ftc2023endorsement} govern advertising-style
disclosure and predate the prose-recommendation surface. MiFID~II
Article~27~\cite{eu2014mifid2} articulates a best-execution analogue in
financial trading---the closest prior regulatory instrument with an
explicit measurement demand---but its measurement apparatus does not
transfer to commission-driven travel recommendation. The DSA
Observatory~\cite{reviglio2024recommender} tracks the shift from ``default +
opt-out'' toward dynamic controls, underscoring that the regulatory target
itself is moving. Adjacent frameworks (EU P2B Regulation 2019/1150 on
platform-to-business transparency~\cite{eu2019p2b}; China's 2022
Algorithm Recommendation Administrative Provisions, CAC Order
No.~9~\cite{china2022algorec}) likewise articulate normative targets
without prescribing quantitative tests. \textit{Across jurisdictions, the pattern
is the same: each instrument names the harm without operationalizing
how to measure it.}

\subsection{What researchers have offered}

Four strands of research tooling bear on LLM-OTA governance. Each has
produced valuable work; each has a structural gap that prevents it from
producing the reading the compliance officer needs.

\paragraph{Travel-planning benchmarks.}
TravelPlanner~\cite{xie2024travelplanner},
TripCraft~\cite{chaudhry2025tripcraft},
TripScore~\cite{qian2025tripscore},
TravelBench~\cite{cheng2025travelbench},
WideHorizon~\cite{yang2025widehorizon}, and
TravelAgent~\cite{chen2024travelagent} measure single-planner task success
against ground-truth itineraries or reward models. They do not model a
commission channel, do not define a welfare rule that governance parameters
can act on, and do not pair a commission message against a counterfactual.

\paragraph{Multi-agent simulations.}
Generative Agents~\cite{park2023generative} initiated a line of
multi-agent social simulation extended by AutoGen~\cite{wu2023autogen},
CAMEL~\cite{li2023camel}, MetaGPT~\cite{hong2024metagpt},
AgentVerse~\cite{chen2024agentverse}, SOTOPIA~\cite{zhou2024sotopia},
GovSim~\cite{piatti2024govsim}, and
Stakeholders~\cite{abdelnabi2024stakeholders}; economically grounded
variants include EconAgent~\cite{li2024econagent},
CompeteAI~\cite{zhao2024competeai}, RecAgent~\cite{wang2025recagent},
Turing Experiments~\cite{aher2023turing}, and Homo
Silicus~\cite{horton2023homo}. Task-competence benchmarks
AgentBench~\cite{liu2024agentbench} and
AgentBoard~\cite{ma2024agentboard} measure within-task progress for a
single agent. None of these parametrizes a welfare rule and sweeps
governance knobs over a paired counterfactual.

\paragraph{Persuasion and sycophancy.}
Sycophancy~\cite{sharma2024sycophancy},
SycEval~\cite{fanous2025syceval}, Durmus et al.~\cite{durmus2024persuasion},
Salvi et al.~\cite{salvi2025persuasiveness}, Hackenburg and
Margetts~\cite{hackenburg2024microtargeting}, Matz et
al.~\cite{matz2024personalized}, Bai et al.~\cite{bai2023aipersuasion},
and Rogiers et al.~\cite{rogiers2024persuasion} quantify whether LLMs can
shift user attitudes, typically in debate, advocacy, or marketing. They
measure user-side preference shift, not platform-induced welfare loss
under a commission objective.

\paragraph{Commercial steering and LLM-mediated commerce.}
A concurrent line of work has begun to measure LLM commercial steering
directly. Salvi et al.~\cite{salvi2026commercial}, in two preregistered
experiments ($N{=}2{,}012$), construct paired counterfactual ebook
selection trials in which one-fifth of products are randomly designated
sponsored: an LLM-mediated interface selects sponsored products at
$61.2\%$ versus $22.4\%$ under traditional search, with most participants
failing to detect the steering; sponsorship label and concealment
instructions are varied as a binary disclosure axis. Bansal et
al.~\cite{bansal2025magentic} release Magentic Marketplace, an
open-source agentic-market simulator that exposes a generic-marketplace
agent to a fixed catalogue of manipulation tactics. These works establish
the phenomenon of LLM commercial steering and a binary-treatment
measurement frame; they do not (i)~target the OTA travel vertical, where
bookings are high-stakes, low-frequency, and explicitly regulated by EU
DSA Article~25~\cite{eu2022dsa}, the FTC Endorsement
Guides~\cite{ftc2023endorsement}, and China's 2022 Algorithm
Recommendation Provisions; (ii)~parametrize commission intensity and
disclosure strictness as continuous orthogonal governance dials acting on
a welfare-rule decision rather than as binary sponsored-versus-not labels;
or (iii)~apply a symmetric producer-side audit that distinguishes
generator-failure modes (over-hedging, template collapse, internal-ID
leakage) from behavioral steering. TourMart positions itself as the
instrument layer that connects the commercial-steering phenomenon to a
per-deployment compliance readout in a regulated commercial vertical,
validated under cross-family replication.

\paragraph{Dark patterns and mechanism design.}
UI-level classification---Mathur et al.~\cite{mathur2019darkpatterns},
Gray et al.~\cite{gray2018darkpatterns, gray2024ontology}, Di Geronimo et
al.~\cite{digeronimo2020uidark}, Luguri and
Strahilevitz~\cite{luguri2021shining}---is post-hoc and captures UI artefacts,
not deployment-time prose under parameter variation.
DarkBench~\cite{kran2025darkbench} extends the taxonomy to LLM outputs;
Ersoy et al.~\cite{ersoy2026darkwebagents} invert the frame, studying LLMs
as \emph{victims} of UI dark patterns. Recommender-fairness frameworks
(FairRec~\cite{patro2020fairrec}) regulate ranking rather than generated
prose. Auction-based aggregation (Duetting et
al.~\cite{duetting2024mechanism}, Soumalias et al.~\cite{soumalias2024rag},
Dubey et al.~\cite{dubey2024summaries}) and collusion analysis (Fish et
al.~\cite{fish2024collusion}) regulate multi-LLM markets through bidding;
they do not instrument a single deployed LLM's prose adjustment relative
to a welfare rule.

\textit{No strand, alone or in combination, produces the sentence a
compliance officer needs.}

\subsection{The gap and our contribution}

Taken together, the prior subsections describe a literature that has
characterized the harm, named it in regulation, and built adjacent tools
for planning, persuasion analysis, dark-pattern detection, and mechanism
design. What none of them supplies---and what a compliance officer, a
board member, or a regulator would need in order to act---is a
\textbf{per-configuration, welfare-rule-anchored, scenario-clustered
behavioral readout, with an external audit on the producer that generated
the prose.}

\noindent Concurrent work has established the
\emph{phenomenon} of LLM commercial steering in adjacent commercial
settings: Salvi et al.~\cite{salvi2026commercial}, in two preregistered
ebook-recommendation experiments ($N{=}2{,}012$), demonstrate substantial
sponsored-product preference under binary disclosure manipulation; Bansal
et al.~\cite{bansal2025magentic} release an open-source agentic-market
simulator. We acknowledge these as the \emph{phenomenon-establishing}
prior art. \textit{TourMart's contribution is not first measurement of
LLM commercial steering---that priority belongs to
Salvi~et~al.~\cite{salvi2026commercial}---but the first
\emph{audit-instrument variant} for this phenomenon in a regulated
commercial vertical, featuring (i)~a continuous $(\lambda, \kappa)$
governance grid acting on a frozen welfare rule (where $\lambda$ is gain
and $\kappa$ saturation cap on a single message-induced welfare
channel), exposing a three-regime structure (attenuated, live
transmission, saturated) that a binary disclosure axis cannot recover
(\S\ref{sec:results}.5); (ii)~a symmetric producer-side six-gate audit
that separates generator-failure modes (template collapse, refusal,
internal-ID leakage) from genuine commercial steering---absent in Salvi
et al.\ and required for any deployment-grade audit; and
(iii)~scenario-clustered grid max-stat correction with cross-family
verification at the audit-gate level (full Llama-OTA paired replay
deferred to future work; see \S\ref{sec:limitations}). The differentiation against \cite{salvi2026commercial}
breaks down along five axes:}

\begin{description}\setlength{\itemsep}{2pt}
  \item[Domain.] Salvi et al.\ study e-book recommendation (consumer
  retail). We study OTA travel, a commercial vertical with explicit
  regulatory anchoring (EU DSA Art.~25, FTC 16~CFR~255, EU P2B 2019/1150,
  China 2022 Algorithm Recommendation Provisions, MiFID~II Art.~27 as
  best-execution analogue).
  \item[Treatment design.] Binary disclosure (sponsored-labeled vs.\
  unlabeled) in Salvi et al.; continuous $(\lambda, \kappa)$ governance
  grid over 36 cells in TourMart, exposing attenuated / live / saturated
  regimes a binary axis cannot recover (\S\ref{sec:results}.5).
  \item[Producer-side audit.] Absent in Salvi et al.\ (single-interface
  measurement). TourMart adds a symmetric six-gate audit (JSON validity,
  bundle coverage, word-count, refusal, unique-message ratio, ID
  leakage) that separates generator failures from commercial steering.
  \item[Cross-family verification.] Salvi et al.\ use one LLM interface.
  TourMart spans two traveler-reader backbones (Qwen-14B-AWQ,
  Llama-3.1-8B), both significant at $\alpha{=}0.05$ (Qwen at primary
  $n{=}143$; Llama via extended-$n$ supplement at $n{=}270$). Producer
  side attempted Mistral and Llama; both surfaced gate failures.
  \item[Mechanism analysis.] Salvi et al.\ report effect-size estimates;
  TourMart adds coefficient-zero attribution decomposing steering across
  four perception channels (fit, trust, risk, urgency).
\end{description}

\section{Problem Formulation}
\label{sec:formulation}

\begin{figure}[t]
\centering
\begin{tikzpicture}[
  node distance=0.7cm and 1.0cm,
  every node/.style={font=\small, align=center},
  box/.style={draw, rectangle, rounded corners=2pt, minimum height=0.7cm, minimum width=2.6cm, fill=gray!8},
  inbox/.style={draw, rectangle, rounded corners=2pt, minimum height=0.7cm, minimum width=2.0cm, fill=blue!10},
  outbox/.style={draw, rectangle, rounded corners=2pt, minimum height=0.7cm, minimum width=2.6cm, fill=green!12},
  audit/.style={draw, rectangle, rounded corners=2pt, minimum height=0.7cm, minimum width=2.6cm, fill=red!12},
  arrow/.style={-{Latex[length=2mm]}, thick}
]
\node[inbox] (m) {Market $M$};
\node[inbox, right=of m] (ota) {OTA $\Theta$};
\node[inbox, right=of ota] (lk) {Dials $(\lambda, \kappa)$};
\node[box, below=of ota] (stim) {Paired stimuli $\mathcal{S}$};
\node[box, below=of stim] (msgs) {$\text{msg}_{\text{orig}}, \text{msg}_{\text{fact}}$};
\node[audit, right=1.0cm of msgs] (gates) {6-gate producer audit};
\node[box, below=of msgs] (perc) {$\phi_{\text{orig}}, \phi_{\text{fact}}$};
\node[box, below=of perc] (rule) {Welfare rule (Eq.~\ref{eq:decision-rule})};
\node[outbox, below=of rule] (out) {$\Delta_{\text{acc}}, p_{\text{McNemar}}, p_{\text{perm}}$};
\node[outbox, right=1.0cm of out] (gout) {gate vector $\mathbf{g}$};
\draw[arrow] (m) -- (stim);
\draw[arrow] (ota) -- (stim);
\draw[arrow] (lk) -- (stim);
\draw[arrow] (stim) -- (msgs);
\draw[arrow] (msgs) -- (perc);
\draw[arrow] (msgs) -- (gates);
\draw[arrow] (perc) -- (rule);
\draw[arrow] (rule) -- (out);
\draw[arrow] (gates) -- (gout);
\end{tikzpicture}
\caption{TourMart audit instrument. Inputs: market $M$, deployed OTA $\Theta$, governance dials $(\lambda, \kappa)$. The OTA produces a commission-aware message and a minimum-disclosure factual template on each paired stimulus; the traveler-reader extracts perception features that feed the welfare rule. Outputs: paired steering delta $\Delta_{\text{acc}}$, significance under exact McNemar and scenario-clustered permutation, plus a producer-side audit gate vector $\mathbf{g}$. See Algorithm~\ref{alg:tourmart} for the full procedure.}
\label{fig:tourmart-pipeline}
\end{figure}
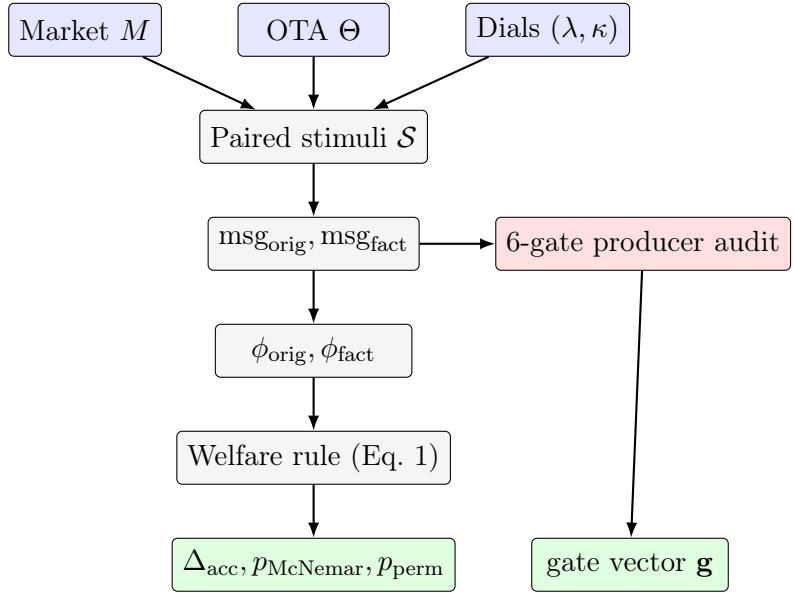

\subsection{The TourMart marketplace}

A small market $M = (T, H, A, B)$ with travelers $T$, hotels $H$,
airlines $A$, and bundles $B$. Each traveler $t \in T$ has a budget
$b_t$, a vibe/archetype with acceptance threshold $\tau_t$, and a utility
$u_t(\beta)$ over bundles $\beta \in B$.

Each bundle $\beta$ has a price $p(\beta)$ decomposed into hotel, airline,
and extras costs. An allocation $\alpha: T \to B \cup \{\emptyset\}$
produces traveler surplus $\sum_t (u_t(\alpha(t)) - p(\alpha(t))) \cdot
\mathbb{1}[\alpha(t) \neq \emptyset]$ and platform revenue equal to the
sum of commission on matched bundles.

\textbf{Oracle welfare ceiling}: computed by MILP over feasible
allocations (details in Section~\ref{sec:setup}).

\subsection{The OTA-traveler interaction}

For each market and condition $c \in \{$commission, satisfaction,
disclosure$\}$:
\begin{enumerate}
  \item The OTA LLM receives traveler profiles plus an observable prior
        (signal strength $w \in [0,1]$) and produces per-traveler
        recommendations with natural-language messages.
  \item Each traveler LLM receives its profile, the OTA's message, and
        the recommended bundle, and extracts 4 perceived features
        $\phi(\beta, \text{msg}) \in [-1,1]^4$ (perceived fit, trust,
        risk, urgency).
  \item A \textbf{deterministic welfare rule} maps those features to an
        acceptance decision.
\end{enumerate}

\subsection{The decision rule with governance parameters}

\begin{equation}
\label{eq:decision-rule}
\begin{aligned}
&\text{acc}(\phi, u_t, p_\beta, b_t, \tau_t;\, \lambda, \kappa) = \\
&\mathbb{1}\Big[ (u_t(\beta) - p_\beta) + \mathrm{clip}(\lambda \vec{c}
\cdot \phi \cdot b_t,\, [\pm\kappa b_t]) \geq \tau_t b_t \Big]
\end{aligned}
\end{equation}

with:
\begin{itemize}
  \item $\lambda$ = coefficient multiplier (Round~20 baseline:
        $\lambda = 1.0$)
  \item $\kappa$ = message adjustment cap as a fraction of budget
        (Round~20: $\kappa = 0.05$)
  \item $\vec{c} = (0.03,\ 0.015,\ -0.025,\ 0.01)$ (fit, trust, risk,
        urgency; signs denote direction of effect on welfare)%
        \footnote{Released evidence files
        \path{results/phase1c_cap_ablation_v4.md} and
        \path{results/phase1c_coef_attribution_v4.md} print
        \texttt{risk=0.025} as a magnitude; the negative sign is absorbed
        into the dot product in Eq.~\ref{eq:decision-rule} above. Both
        conventions describe the same welfare rule; higher perceived risk
        reduces welfare.}
  \item Hard floor: if $u_t(\beta) - p_\beta < -0.10 \cdot b_t$, reject
        regardless.
\end{itemize}

$\boldsymbol{\lambda}$ \textbf{and} $\boldsymbol{\kappa}$ \textbf{are the
governance parameters} we vary in the phase diagram.

\begin{figure}[t]
\centering
\begin{minipage}{0.96\linewidth}
\hrule
\vspace{2pt}
\textbf{Algorithm 1: TourMart Audit Instrument}\hfill\textit{(\refstepcounter{equation}label \theequation\label{alg:tourmart})}
\vspace{2pt}
\hrule
\vspace{4pt}

\textbf{Input:} Market $M = (T, H, A, B)$; deployed OTA backbone $\Theta$; traveler-reader $\Pi$; governance dials $(\lambda, \kappa)$; near-threshold sample $\mathcal{S}$ with $|\mathcal{S}|=n$.

\textbf{Output:} Audit reading $\mathcal{R} = (\Delta_{\text{acc}}, p_{\text{McNemar}}, p_{\text{perm}}, \mathbf{g})$.

\vspace{4pt}
\hrule
\vspace{4pt}

\begin{tabbing}
\hspace{2.0em}\=\hspace{1.5em}\=\kill
1.\> \textbf{for} each paired stimulus $(t, \beta) \in \mathcal{S}$ \textbf{do} \\
2.\> \> $\text{msg}_{\text{orig}} \gets \Theta(t, \beta;\, \text{commission-signal})$ \\
3.\> \> $\text{msg}_{\text{fact}} \gets \text{factual\_template}(t, \beta)$ \\
4.\> \> $\phi_{\text{orig}} \gets \Pi(t, \beta, \text{msg}_{\text{orig}})$;\quad $\phi_{\text{fact}} \gets \Pi(t, \beta, \text{msg}_{\text{fact}})$ \\
5.\> \> $a_{\text{orig}}[t,\beta] \gets \mathrm{acc}(\phi_{\text{orig}}, u_t, p_\beta, b_t, \tau_t;\, \lambda, \kappa)$\quad \textit{(Eq.~\ref{eq:decision-rule})} \\
6.\> \> $a_{\text{fact}}[t,\beta] \gets \mathrm{acc}(\phi_{\text{fact}}, u_t, p_\beta, b_t, \tau_t;\, \lambda, \kappa)$ \\
7.\> \textbf{end for} \\
8.\> $\Delta_{\text{acc}} \gets \tfrac{1}{n} \sum_{(t,\beta) \in \mathcal{S}} ( a_{\text{orig}}[t,\beta] - a_{\text{fact}}[t,\beta] )$ \\
9.\> $(b, c) \gets$ count discordant pairs where $a_{\text{orig}} \ne a_{\text{fact}}$ \\
10.\> $p_{\text{McNemar}} \gets \mathrm{ExactMcNemar}(b, c)$ \\
11.\> $p_{\text{perm}} \gets \mathrm{ScenarioClusteredPermutation}(\mathcal{S},\, n_{\text{perm}}{=}1000)$ \\
12.\> $\mathbf{g} \gets \mathrm{SixGateAudit}(\{\text{msg}_{\text{orig}}\}_{(t,\beta)\in\mathcal{S}})$ \\
13.\> \textbf{return} $\mathcal{R} = (\Delta_{\text{acc}}, p_{\text{McNemar}}, p_{\text{perm}}, \mathbf{g})$
\end{tabbing}

\vspace{2pt}
\hrule
\end{minipage}
\caption{TourMart audit procedure: paired counterfactual generation under
fixed traveler and bundle, deterministic welfare-rule scoring, exact and
permutation significance, and producer-side gate audit.}
\label{alg:tourmart-fig}
\end{figure}

\subsection{Estimand}

The paired acceptance delta
\[
\Delta_{\text{acc}} = \frac{1}{n} \sum_{(m, t, \beta) \in \mathcal{S}}
\big[ \text{acc}(\phi^{\text{orig}}, \cdot) - \text{acc}(\phi^{\text{fact}},
\cdot) \big]
\]
over the sampled near-threshold stimuli $\mathcal{S}$, where
$\phi^{\text{orig}}$ is extracted from the OTA's free-text commission
message and $\phi^{\text{fact}}$ from a minimum-disclosure factual
template.

\textbf{Pre-registered primary endpoint}: $\Delta_{\text{acc}}$ at
deployed parameters $(\lambda = 1, \kappa = 0.05)$ for each traveler
backbone, tested by exact McNemar on discordant pairs.

\section{The Commission-Messaging Treatment}
\label{sec:treatment}

\subsection{OTA prompt}

The OTA agent receives:
\begin{itemize}
  \item Market state (bundles, travelers, budgets, capacities).
  \item Observable preference prior (signal-weighted).
  \item Condition-specific objective block:
    \begin{itemize}
      \item \textbf{Commission}: ``maximize platform commission''
      \item \textbf{Satisfaction}: ``maximize traveler welfare''
      \item \textbf{Disclosure-compliant}: mandatory disclosures block
    \end{itemize}
\end{itemize}
It outputs a JSON payload with \texttt{decision\_table} and
\texttt{recommendations} (\texttt{bundle\_id}, \texttt{message},
\texttt{disclosures}).

\subsection{Factual template}

Minimum-disclosure floor:
\begin{quote}
``Recommend $\beta$. Total price: $p$. Commission rate: $r$\%. Extras:
$E$.''
\end{quote}
This represents the operational counterfactual---what a regulator could
mandate as a minimum-factual-disclosure floor. It is \textbf{not} a
length-matched persuasion-isolation control; that is a different estimand
(future work).

\subsection{Traveler hybrid architecture}

Each traveler LLM uses guided JSON to produce \texttt{\{perceived\_fit\_delta:
float, perceived\_risk: float, trust\_score: float, urgency\_felt: float,
reasoning: str\}}. A deterministic welfare rule then evaluates acceptance.

\section{Experimental Setup}
\label{sec:setup}

\subsection{Models (locked Round~20)}

\begin{itemize}
  \item \textbf{OTA backbone}: Qwen2.5-7B-Instruct, bf16, vLLM with
        guided JSON and per-market \path{bundle_id} enum,
        \path{max_tokens=2500}.
  \item \textbf{Traveler backbones}: Qwen2.5-14B-Instruct-AWQ (float16,
        \path{awq_marlin}); Llama-3.1-8B-Instruct (bfloat16).
  \item \textbf{Welfare rule}: frozen Round~20 coefficients;
        $\text{MSG\_ADJ\_CAP} = 0.05 \cdot \text{budget}$;
        $\text{baseline\_floor} = -0.10 \cdot \text{budget}$.
\end{itemize}

\subsection{Markets and scenarios}

\begin{itemize}
  \item Small-market generator \texttt{generate\_small\_market(seed,
        regime)} with 3--5 hotels, 2--3 airlines, 3--6 travelers, 6--10
        bundles.
  \item \textbf{Scenario pool}: 75 loose regime + 75 tight regime, 3
        episode seeds $\rightarrow$ 450 per regime, 900 total (v4 scale).
  \item Conditions: commission / satisfaction / disclosure\_compliant
        (only commission used for Phase~1c replay).
  \item Signal weight $w \in \{0.25, 0.5, 0.75\}$ crossed with regime.
\end{itemize}

\subsection{Near-threshold window (diagnostic)}

\[
\text{baseline\_surplus} \in [\tau_t b_t - 0.10 b_t,\ \tau_t b_t + 0.05
b_t]
\]
Stratified sampling per $(w, \text{regime})$, target 100 per stratum,
seed $= 42$.

\subsection{Paired sampling}

For each sampled recommendation, the traveler LLM evaluates two variants:
\begin{itemize}
  \item \textbf{Original}: OTA's full commission-condition message.
  \item \textbf{Factual}: minimum-disclosure template.
\end{itemize}
Paired by the 3-tuple key $(\text{scenario}, \text{traveler},
\text{bundle})$; the diagnostic-window filter plus this pairing yields
143 paired stimuli across 88 scenarios (v4).%
\footnote{The stratified sampling can produce multiple
$(\text{signal\_wt}, \text{episode\_seed})$ realizations for the same
$(\text{scenario}, \text{traveler}, \text{bundle})$ 3-tuple; under the
3-tuple pair-up, the last realization per 3-tuple is retained and the
earlier ones are collapsed. A strict full-identity 5-tuple
$(\text{scenario}, w, \text{seed}, \text{traveler}, \text{bundle})$
re-pairing on the same raw data yields $n{=}409$ pairs and shifts the
deployed- and peak-point effect sizes (see reproduction artifact
\path{scripts/reproduce_permutation.py} and the sibling 5-tuple variant).
The headline numbers reported in Table~\ref{tab:1}, the abstract, and the
Results section are the 3-tuple analysis; we release the
episode-seed-annotated raw log
(\path{results/phase1c_*_diag_v4_report.with_episode_seed.raw.jsonl}) so
external verifiers can reproduce either pairing.}

\subsection{Statistical plan (pre-registered)}

\begin{enumerate}
  \item \textbf{Primary (per arm)}: McNemar exact on discordant pairs at
        deployed $(\lambda = 1, \kappa = 0.05)$.
  \item \textbf{2D phase diagram (secondary-but-predeclared)}: $\lambda
        \in \{1, 2, 3, 5, 10, 20\} \times \kappa \in \{1\%, 2.5\%, 5\%,
        10\%, 20\%, 100\%\}$: 36 cells.
  \item \textbf{Family-wise correction}: grid-level max-stat permutation
        with scenario as cluster unit (1000 permutations).
  \item \textbf{Exploratory feature-level}: Holm-corrected paired
        Wilcoxon on the 4 feature deltas.
\end{enumerate}

\subsection{Validity gates}

\begin{itemize}
  \item Parse success $\geq 95\%$.
  \item Factual acceptance not at ceiling ($< 98\%$).
  \item Near-threshold window mechanically flippable.
\end{itemize}

\subsection{OTA-backbone audit (symmetric stimulus)}

Our main results use Qwen-7B-Instruct as the OTA backbone. As a
robustness diagnostic, we audited Llama-3.1-8B-Instruct as a candidate
OTA under the same SYSTEM\_PROMPT and vLLM guided-JSON configuration. The
audit applies six gates symmetrically to both candidate and comparator
msgcap: (i)~JSON validity $\geq 85\%$, (ii)~bundle-id coverage on success
recommendations $\geq 80\%$, (iii)~message word-count median $\in [10,
200]$, (iv)~refusal/hedging rate $\leq 20\%$ under a \textbf{regex-based
hardened refusal classifier} (regex v2---11 additions over v1:
\texttt{unfortunately}, \texttt{couldn't find}, \texttt{could not find},
\texttt{unable to find}, \texttt{cannot find}, \texttt{did not find},
\texttt{no suitable}, \texttt{not suitable}, \texttt{no bundle matches},
\texttt{does not match your}, \texttt{doesn't match your}), (v)~unique-message
ratio on success-only $\geq 30\%$, (vi)~internal-ID leakage (regex
\texttt{t$\backslash$d+} or \texttt{b$\backslash$d+} appearing in the
message body) $\leq 20\%$. The classifier is a regex by implementation
and a refusal decision rule by function; we use ``regex-based refusal
classifier'' consistently throughout. The full protocol,
pre-registration-in-spirit guardrail, and all four audit markdown outputs
are reported in Appendix~\ref{app:sensitivity}.3 and in
\texttt{refine-logs/PHASE1C\_SECTION\_5\_3\_FINDING.md}.

\section{Results --- Cross-family Phase Diagram (Qwen-OTA, n=143)}
\label{sec:results}

\subsection{Primary endpoint at deployed parameters}

\begin{table*}[t]
\centering
\caption{McNemar paired acceptance test at deployed $(\lambda=1,
\kappa=0.05)$. $b/c$: discordant$_+$/discordant$_-$.}
\label{tab:1}
\small
\setlength{\tabcolsep}{4pt}
\begin{tabular}{@{}lrcccrc@{}}
\toprule
Arm & $n$ & $b/c$ & orig\,/\,fact acc & RD & 95\% CI & McNemar $p$ \\
\midrule
Qwen-14B-AWQ & 143 & \textbf{12/1} & 56.64\,/\,48.95\% & \textbf{+7.69pp}
  & [+2.88,\,+13.24] & \textbf{0.0034} \\
Llama-3.1-8B & 143 & 5/0 & 64.34\,/\,60.84\% & +3.50pp
  & [+0.71,\,+6.58] & 0.0625 \\
\bottomrule
\end{tabular}
\end{table*}

\textbf{The Qwen-reader arm is significantly steered by commission messages
at deployed parameters.} The Llama-reader RD point estimate is positive
and its estimation interval excludes zero, but under our pre-registered
two-sided exact McNemar decision rule even the most one-sided 5/0 split
among five discordant pairs yields the minimum achievable $p$-value
$p = 2 \cdot (1/2)^5 = 0.0625$; we therefore report the deployed Llama
effect as not significant at $\alpha = 0.05$ despite the positive
estimation interval. The RD estimation CI and the exact McNemar test
address different inferential objects (RD magnitude vs.\ a discordant-pair
sign test), and the two disagree when the discordant count $b{+}c$ is
small; we treat the pre-registered exact test as the decision rule and
flag this as an inference-method sensitivity in Section~\ref{sec:limitations}.

\paragraph{B.1 supplemental robustness check (Llama extended-$n$).}
The Llama-reader $b{+}c{=}5$ at deployed parameters places the
pre-registered exact McNemar test at its minimum-achievable
$p{=}0.0625$, leaving the under-power explanation
indistinguishable from a true null. To separate these we ran a
post-submission supplemental replay of the Llama-3.1-8B reader on the
same Qwen-7B-OTA \texttt{msgcap\_v4} source, widening the near-threshold
selection from primary to diagnostic
($[\tau b - 0.10 b, \tau b + 0.05 b]$, per-stratum capped at 45) to lift
$n$ from 143 to \textbf{270}. The extended sample yields
\textbf{$+2.96$pp paired RD}, \textbf{$b/c{=}8/0$}, \textbf{exact
two-sided $p{=}0.0078$} (the achievable floor at $b{+}c{=}8$),
$95\%$~CI~$[+0.65, +5.95]$pp---\emph{cross-family confirmation at
$\alpha{=}0.05$}. The extended-$n$ point estimate is slightly attenuated
relative to the primary-window $+3.50$pp, consistent with the diagnostic
window pulling in pairs that are mechanically less flippable. The
producer-side six-gate audit on the same upstream msgcap reproduces the
known Qwen over-hedging failure mode ($55.9\%$ refusal) and passes the
five other gates. The supplemental replay artifacts (raw paired JSONL,
selection manifest, verdict markdown) are released as
\texttt{B1\_VERDICT\_BUNDLE.tar.gz} alongside the reproducibility package.

\paragraph{Clip-rate diagnostic.} The clip operation in
Eq.~\ref{eq:decision-rule} binds whenever $|\lambda \vec{c}\cdot\phi| >
\kappa$. Across $n = 818$ paired stimuli (409 original $+$ 409 factual
per arm), empirical $|\vec{c}\cdot\phi|$ has mean $0.016$ on the Qwen
arm and $0.034$ on the Llama arm, with maxima of $0.045$ and $0.048$
respectively---both \emph{below} the deployed $\kappa = 0.05$. At
deployed $(\lambda{=}1, \kappa{=}0.05)$ the cap therefore binds on
$\mathbf{0/818}$ ($\mathbf{0.0\%}$) stimuli on both arms: $\kappa$ is
\emph{prophylactic, not active}. The full grid-level clip-binding map (computed from the released
paired-output JSON; analysis snippet in
\path{reproducibility/scripts}) shows $\kappa$ activating as $\lambda$
scales: at $(\lambda{=}1, \kappa{=}0.025)$ the Qwen arm has $7.5\%$
clip-binding while Llama reaches $88.9\%$, and at $(\lambda{=}3,
\kappa{=}0.10)$ Qwen sits at $7.2\%$ while Llama is already at
$68.1\%$---a qualitative differential bite pattern across $(\lambda,
\kappa)$ cells that a binary disclosure axis cannot expose. The deployed
point is therefore the \emph{least-saturated} end of the grid; the
$+7.69$pp Qwen reading is the unsaturated linear effect of
message-induced perception, and $\kappa$ becomes a governance lever only
when $\lambda$ is scaled or when the perception extractor produces
near-envelope $|\vec{c}\cdot\phi|$.

\subsection{Heterogeneity by signal strength}

\begin{table}[t]
\centering
\caption{Qwen heterogeneity. $w=0.75$ shows a strong main effect.}
\label{tab:2}
\small
\begin{tabular}{lrrrr}
\toprule
signal\_wt & $n$ & RD & disc$_+$/disc$_-$ & McNemar $p$ \\
\midrule
0.25 & 46 & +6.52pp & 3/0 & 0.25 \\
0.5 & 42 & 0.00pp & 1/1 & 1.00 \\
\textbf{0.75} & \textbf{55} & \textbf{+14.55pp} & \textbf{8/0}
  & \textbf{0.0078} $\star$ \\
\bottomrule
\end{tabular}
\end{table}

Higher signal-strength (more targeted recommendations) amplifies
commission-steering. This is expected if steering exploits
preference-matched framings.

\subsection{The 2D phase diagram}

\begin{figure}[t]
\centering
\includegraphics[width=\linewidth]{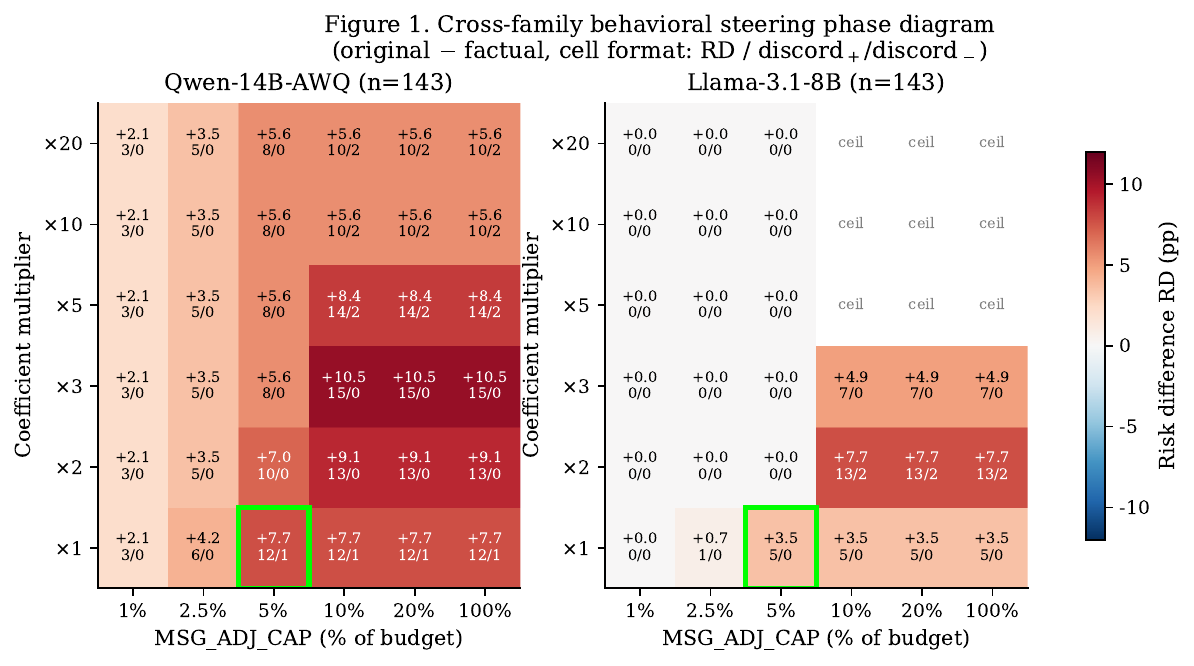}
\caption{Behavioral heatmap: paired RD across the $\lambda \times \kappa$
grid for both traveler-reader arms (Qwen-14B-AWQ top, Llama-3.1-8B
bottom). Green box marks the deployed Round~20 cell $(\lambda=1,
\kappa=0.05)$.}
\label{fig:1}
\end{figure}

Key features (Figure~\ref{fig:1}):
\begin{itemize}
  \item \textbf{Qwen}: monotone-ish increasing in $(\lambda, \kappa)$ up
        to $(\lambda = 3, \kappa = 0.10)$, peak \textbf{+10.5pp, 15/0}.
        Above $\lambda = 5$, mixed discord (factual begins to ceiling on
        some pairs).
  \item \textbf{Llama}: non-monotone. Deployed cell marginal ($+3.5$pp,
        5/0). Live-zone peak at $(\lambda = 2, \kappa = 0.10)$:
        \textbf{+7.7pp, 13/2}. Above $\lambda = 3$ with $\kappa \geq
        0.10$, the factual template itself drives 100\% acceptance
        (ceiling)---this is the \textit{disclosure-saturated} regime and
        is validity-excluded.
\end{itemize}

\subsection{Family-wise corrected significance}

\begin{table}[t]
\centering
\caption{Grid-level max-stat permutation (primary correction).}
\label{tab:3}
\footnotesize
\setlength{\tabcolsep}{4pt}
\begin{tabular}{@{}lrrr@{}}
\toprule
Arm & obs max RD & Pair $p$ & \textbf{Cluster $p$ (PRIMARY)} \\
\midrule
Qwen-14B & +10.49pp @($\times 3$, 10\%) & $<0.001$ & $\mathbf{<0.001}$ \\
Llama-8B & +7.69pp @($\times 2$, 10\%) & 0.007 & \textbf{0.008} \\
\bottomrule
\end{tabular}
\end{table}

\textbf{Procedure}: the unit of exchangeability is the scenario cluster
(88 unique scenarios $\times$ up to 3 episode seeds, 143 pairs total).
For each of 1000 permutations we randomly flip orig$\leftrightarrow$fact
labels within each scenario cluster (preserving cluster membership),
recompute the 36-cell RD grid, and record the max RD across cells. The
cluster-level $p$ is the fraction of permutations whose max RD matches or
exceeds the observed max RD. This procedure corrects for (a)~the 36
grid-cell multiplicity and (b)~within-scenario correlation across episode
seeds and grid cells.

\textbf{Both arms pass family-wise correction under scenario-clustered
permutation.} The live transmission region $(\lambda \in [2, 3], \kappa
\in [5\%, 10\%])$ is not a spurious peak; it is a robust parametric
neighborhood.

\subsection{Three regimes emerge}

\begin{enumerate}
  \item \textbf{Attenuated zone} at $\lambda \leq 1, \kappa \leq 5\%$
        for Llama: the rule admits the deployed $+3.5$pp reading but does
        not cross exact-McNemar significance at $b{+}c{=}5$.
  \item \textbf{Live transmission region} at $\lambda \in [2, 3], \kappa
        \in [5\%, 10\%]$: both arms produce significant behavioral flips
        with perceptual shifts of $\Delta_{\text{fit}} \approx +0.17$.
  \item \textbf{Saturated zone} at $\lambda \geq 5, \kappa \geq 10\%$
        for Llama: even the factual template triggers 100\% acceptance
        because Llama perceives minimum-disclosure prose as substantively
        positive.
\end{enumerate}

\noindent The three-regime structure is what a binary
incentive$\times$disclosure ablation cannot recover: a single low-vs-high
$\kappa$ contrast at fixed $\lambda$ cannot distinguish a live
transmission region (where governance dials matter) from a saturated
regime (where the disclosure floor itself ceilings) on the Llama arm, and
cannot localize the live-region peak at $(\lambda{=}2, \kappa{=}0.10)$
versus the deployed point. The continuous grid is therefore an
information-bearing instrument over a binary design, not a denser
re-tabulation.

\section{Mechanism: What Drives Transmission?}
\label{sec:mechanism}

\subsection{Coefficient attribution}

\begin{figure}[t]
\centering
\includegraphics[width=\linewidth]{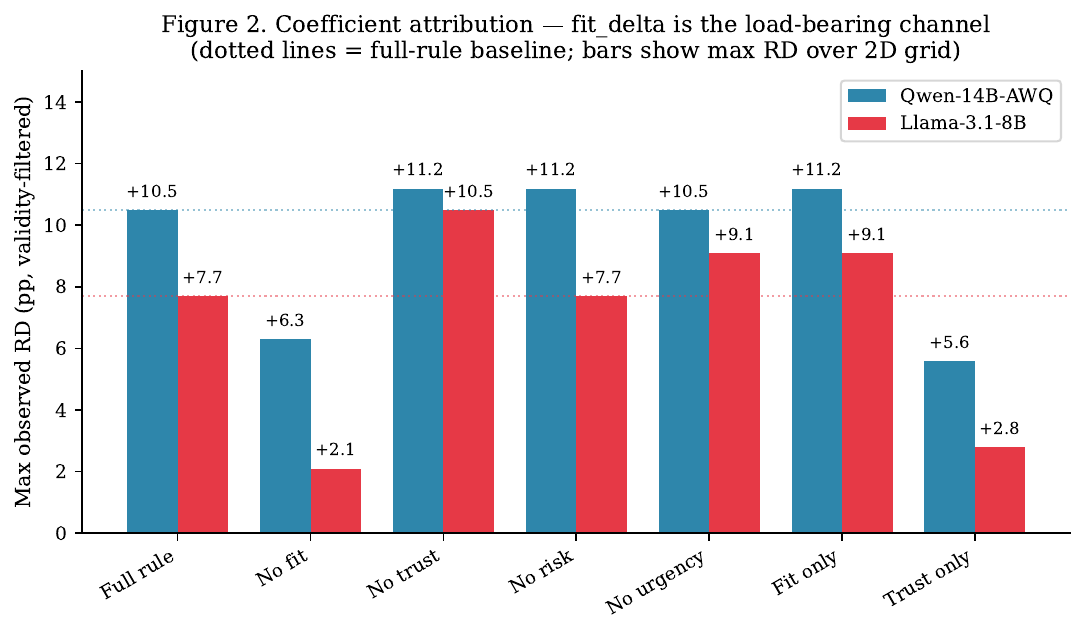}
\caption{Coefficient-zero attribution: max RD after setting each of the
four rule coefficients to zero, recomputed over the 2D grid for each
traveler-reader arm.}
\label{fig:2}
\end{figure}

Figure~\ref{fig:2} reports a rule-level attribution holding extracted
perceptions fixed; \textbf{this is not a causal mediation estimate.} We
do not intervene on the perception extractor or on the underlying message
content---only on the downstream deterministic rule---so the following
should be read as ``which coefficient channel carries the observed RD
under the frozen rule,'' not ``which perceptual construct causally
mediates steering.''

\begin{itemize}
  \item \textbf{Removing \texttt{perceived\_fit\_delta}}: Qwen max RD
        collapses from $+10.5$ to $+6.3$pp; Llama max RD collapses from
        $+7.7$ to \textbf{$+2.1$pp} ($\approx 3.7\times$ reduction).
  \item \textbf{Removing \texttt{trust\_score}}: both arms INCREASE in
        max RD (Qwen $+11.2$, Llama $+10.5$). This is because Llama
        reads the factual template as already high-trust; including trust
        in the rule raises the factual baseline's adjustment and slightly
        \textit{compresses} the differential.
  \item \textbf{Removing \texttt{risk}}: small shift ($+0.7$pp on Qwen,
        unchanged on Llama)---much less than the $\sim 5.6$pp fit-channel
        collapse.
  \item \textbf{Removing \texttt{urgency}}: small shift ($+1.4$pp on
        Llama, unchanged on Qwen)---also small relative to the fit
        channel.
\end{itemize}

\subsection{Baseline perception channels differ by traveler-reader backbone}

Llama's factual-variant baseline perception on the paired $n=143$ stimuli
(mean fit\_delta $= +0.598$, mean trust $= +0.824$) is descriptively much
higher than Qwen's on the same minimum-disclosure template (fit $= +0.010$,
trust $= +0.873$). This is a descriptive backbone difference, not a causal
explanation of transmission. But it matters for governance: a
minimum-disclosure template is not a perceptual zero-point; its position
varies by traveler-reader backbone. Consequently, identical regulatory
disclosures may produce different welfare outcomes across user populations
served by different LLM interfaces. Per-variant (orig vs.\ fact) paired
means on the $n{=}143$ stimuli are reported in the shipped Phase~1c
calibration files
(\path{results/phase1c_qwen14b_awq_diag_v4_report.md} and
\path{results/phase1c_llama31_8b_diag_v4_report.md}); pooled per-feature
distributions (both variants combined, all 818 calls) are in
Appendix~\ref{app:gates}.3, and agree with the same cross-reader pattern.

\subsection{Evidence trajectory across scale-up}

\begin{figure}[t]
\centering
\includegraphics[width=\linewidth]{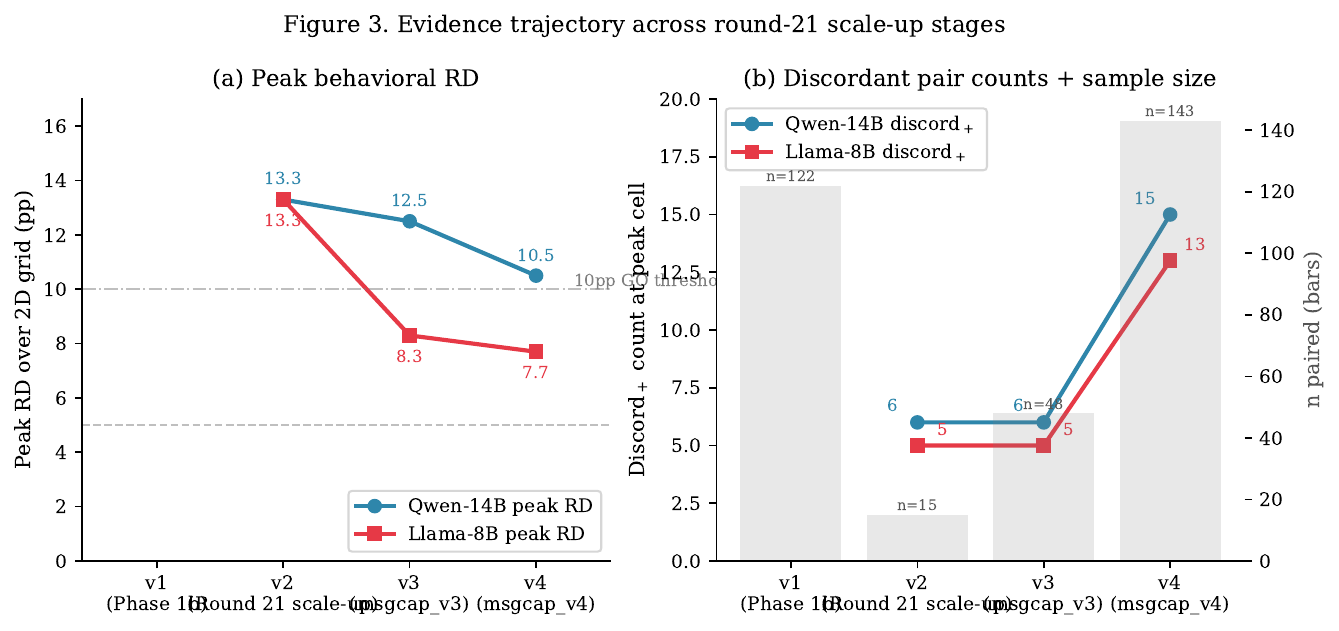}
\caption{Evidence trajectory across v1 ($n=122$, feature-only), v2
($n=15$), v3 ($n=48$), v4 ($n=143$). The effect size stabilizes at
$\sim 10$pp Qwen / $\sim 8$pp Llama as sample size grows, with discord
counts scaling roughly linearly.}
\label{fig:3}
\end{figure}

Figure~\ref{fig:3} reports peak RD and discordant counts across the
pilot-to-v4 runs.

\section{Discussion}
\label{sec:discussion}

\subsection{What TourMart is (and is not)}

\textbf{Is}: an applied intelligent-system audit instrument for governance mechanisms
over the LLM-OTA $\rightarrow$ traveler perception $\rightarrow$
welfare-rule pipeline. A regulator or platform can use TourMart to answer
``how sensitive is this decision rule to commission-maximizing
persuasion?''

\textbf{Is not}: direct evidence that real Booking.com users would be
steered by $+10.5$pp. The estimand is transmission through a frozen
welfare rule in a simulator, across multiple traveler backbones.

\subsection{Governance implications}

The three-regime structure (attenuated / live / saturated) gives platform
operators and regulators a parametric map. Deployed rules can sit safely
in the attenuated zone (our Round~20 parameters), but the boundaries are
narrow---a single coefficient tightness $\times$ cap loosening can move a
rule from safe to steerable. This argues for \textbf{governance parameter
transparency} (e.g., platforms disclosing the effective $\lambda, \kappa$
in auditable form).

\subsection{Disclosure heterogeneity by user model}

Llama's ``disclosure-saturated zone'' (factual text triggers 100\%
acceptance) is a cautionary calibration finding: \textbf{``neutral
disclosure'' is not neutral}. Its effect on user choice depends on the
user-side model/representation. Benchmarks that evaluate disclosure floors
must report \textit{paired} orig-vs-factual effects, not \textit{absolute}
factual acceptance. This `disclosure-saturated' pattern---where mandated
transparency saturates rather than informs the recipient---replicates in
the LLM-OTA setting a finding that legal and behavioral-economics work
has documented for over a decade in human-decision settings
(\cite{benshahar2014disclosure,loewenstein2014disclosure}); our
contribution here is to show the same failure mode reproduces when
disclosure is interpreted by an LLM rather than a human reader, and to
locate the threshold ($\lambda \geq 5, \kappa \geq 10\%$ on Llama) at
which it activates in our setting. A parallel heterogeneity appears on the
\textit{producer} side (Section~\ref{sec:setup}.7,
Appendix~\ref{app:sensitivity}.3): the same default SYSTEM\_PROMPT under
vLLM guided JSON elicits two qualitatively different failure profiles from
Qwen-7B and Llama-3.1-8B---Qwen over-hedges (55.9\% of commission
responses are refusals/hedges under the regex-based hardened refusal
classifier), Llama template-collapses (median 6 words, 80.9--84.6\%
internal-ID leakage). The Llama failure is text-surface repairable by an
explicit 2-sentence / 15--40-word style-constraint suffix
(Section~\ref{sec:setup}.7 audit gates all pass at seeds 42 and 43,
Appendix~\ref{app:sensitivity}.3.b--c); a parallel style-suffix repair
for Qwen's hedging was not evaluated in this submission and is flagged as
future work. The producer-side heterogeneity suggests that OTA-audit
regimes must specify which failure modes are monitored rather than
assuming a single text-quality metric suffices.

\section{Limitations}
\label{sec:limitations}

\begingroup\sloppy
\begin{enumerate}
  \item \textbf{Simulator scope}: the welfare rule is stylized; real
        travelers integrate information across multiple touches
        (comparisons, reviews, price history) and engage in multi-turn
        refinement and feedback loops that TourMart does not model.
  \item \textbf{Single OTA family for primary results}: our Phase~1c
        paired replay uses Qwen-7B-Instruct as the OTA backbone. We
        attempted two additional families. Mistral-7B-Instruct-v0.3
        failed the 85\% JSON-validity gate under our vLLM
        structured-output config and was deferred. Llama-3.1-8B-Instruct
        passed JSON validity and bundle-id coverage but failed three
        content gates under the default SYSTEM\_PROMPT (word-count median
        $=6$, unique-success ratio $=4.0\%$--$12.5\%$, internal-ID
        leakage $=80.9\%$--$84.6\%$), a template-collapse pattern that
        replicated across two sampling seeds (42, 43) and was not
        explained by batch size (Section~\ref{sec:setup}.7,
        Appendix~\ref{app:sensitivity}.3). The collapse was repaired
        at the audit-gate level by an explicit 2-sentence / 15--40-word
        style-constraint suffix appended to the SYSTEM\_PROMPT, but the
        full 143-episode paired steering replay under the repaired
        prompt was not run in this submission. We therefore frame the
        Llama-OTA result as a \textbf{prompt-sensitivity finding for the
        OTA-language surface}, not a cross-family robustness claim for
        TourMart's main steering estimands.
  \item \textbf{Two traveler families only}: Gemma-2-9B, GPT-4o-mini, and
        other traveler backbones are not evaluated here.
  \item \textbf{Near-threshold window selection and conditional estimand}:
        we test the region where the welfare rule is mechanically
        flippable. Pairs with baseline well above or below $\tau \cdot
        b_t$ are deterministically accepted/rejected; the phase diagram
        characterizes the marginal regime. More fundamentally, the paired
        replay identifies message transmission \textbf{conditional on the
        OTA's recommended bundle set}; it does not estimate total platform
        steering from the joint bundle-selection-plus-message policy.
  \item \textbf{LLM-extracted perception features and ``bundle fit''
        construct validity}: the traveler's perception is itself generated
        by an LLM. This is a fundamental constraint of LLM-agent
        simulators; we do not claim population-level generalization to
        humans. It also leaves open a construct-validity concern for
        Section~\ref{sec:mechanism}: the \texttt{perceived\_fit\_delta}
        channel may, in part, be the LLM's default encoding of whatever
        persuasive signal it received into the coefficient with the
        largest downstream weight. Human-annotated fit validation on a
        stratified subset of messages and a fit-language-removal
        counterfactual (planned as future work) are the strongest
        mitigations.
  \item \textbf{Small-discordance and small-stratum power}: headline
        effects rest on small numbers of discordant pairs (Qwen 12/1,
        Llama 5/0 out of 143 each) and small per-signal-strength strata
        ($n = 42$--$55$). Exact McNemar is conservative at small
        $b{+}c$; in particular the Llama-reader deployed $p=0.0625$ is
        the smallest possible two-sided exact $p$-value at $b{+}c{=}5$.
        The B.1 supplemental extended-$n$ replay
        (\S\ref{sec:results}.1) directly addresses this caveat for the
        Llama arm: at $n{=}270$ in the diagnostic window, $b{+}c{=}8$
        clears the floor and yields exact $p{=}0.008$, ruling out
        under-power as the explanation for the deployed $p{=}0.0625$.
        Grid-level max-RD peaks (Qwen 15/0, Llama 13/2) are also
        small-count effects. We treat the 88 unique scenario clusters as
        the exchangeable unit in the permutation test; shared
        bundle-generator seeds, shared traveler-archetype templates, and
        shared market-generator draws across episode seeds may induce
        residual dependence that 88 clusters underweight, which we
        cannot fully rule out without independently generated scenarios.
        We report exact tests and estimation intervals separately and
        use the pre-registered exact test as the significance decision;
        subgroup claims (e.g., $w{=}0.75$ McNemar $p{=}0.0078$) are
        reported as exploratory within the corrected primary test.
  \item \textbf{Audit-gate operational caveats}: excluding Mistral-7B
        and default-prompt Llama-3.1-8B from the paired steering replay
        conditions the audit instrument on the OTA passing the six-gate
        audit. This is the correct operational choice for a
        text-in-the-loop instrument but limits claims about OTA models
        ``in the wild'' whose deployment pipeline may not enforce the
        same gates. The refusal classifier is regex-based: a triage
        signal, not a semantic quality guarantee, with both false
        positives (a legitimate refusal that uses a non-matching phrase
        is missed) and false negatives (a recommendation that happens
        to contain a hedging substring is flagged). Two of the six
        audit-gate thresholds were widened after the Qwen baseline
        failure profile was observed---the word-count median window
        from $[30, 200]$ to $[10, 200]$ and the refusal cutoff from
        $10\%$ to $20\%$ (Appendix~D.3); to our knowledge no widely-cited
        dark-pattern or refusal benchmark fixes a $20\%$ refusal cutoff,
        so the choice is a TourMart-internal documentation threshold.
        The widening did not move any arm from fail to pass: Qwen still
        fails the looser $20\%$ refusal gate at $55.9\%$, and Llama
        still fails three content gates under the default prompt.
  \item \textbf{Pre-registered decision rule = frozen}: the Round~20
        coefficients were chosen deliberately to sit in the attenuated
        regime. The phase diagram shows what happens if the rule drifts;
        it does not prove any specific $(\lambda, \kappa)$ is the
        ``right'' choice, and we offer no external calibration for the
        deployed point beyond the validity gates.
  \item \textbf{$\lambda$--$\vec{c}$ identification}: under the deployed
        coefficient vector
        $\vec{c} = (0.03,\, 0.015,\, {-}0.025,\, 0.01)$,
        the gain dial $\lambda$ is identified as a multiplicative
        deviation from the baseline rule. Under a free $\vec{c}$ the
        composition $\lambda\vec{c}$ is only identified up to a positive
        scalar, so $\lambda$ alone would be a normalization artifact;
        we therefore treat the deployed $\vec{c}$ as part of the
        welfare-rule \emph{specification}, not a free hyperparameter.
        Audits of platforms with a different welfare rule must restate
        their $\vec{c}$ to make $\lambda$-readings comparable across
        deployments.
\end{enumerate}
\endgroup

\section{Conclusion}
\label{sec:conclusion}

We introduced TourMart, a parametric audit instrument for measuring LLM-OTA
commission-steering under disclosure governance. With $n=143$ paired
near-threshold same-bundle replays across two traveler-reader backbones,
we demonstrated scenario-clustered family-wise-corrected behavioral
steering ($+10.49$pp Qwen-reader, $+7.69$pp Llama-reader peak RD),
characterized a three-regime transmission phase diagram, and, under
coefficient-zero attribution holding extracted perceptions fixed, found
that the perceived-fit channel carries most of the observed max-RD signal.
TourMart's contribution is operational: it lets regulators and platform
operators characterize the steering susceptibility of a proposed decision
rule, and it surfaces the non-trivial fact that \textit{minimum-disclosure
templates are not perceptual zero points} across user-side model families.
We release the audit instrument, benchmark scenarios, welfare-rule ablation code, four
symmetric OTA-text audit reports, and all paired raw traveler outputs to
support reproduction and extension.

\section*{Ethics Statement}

\textbf{Dual-use considerations.} TourMart is a governance-audit
benchmark: it measures when commission-maximizing OTA messages transmit
into traveler-agent decisions, and which governance parameters attenuate
transmission. The same instrument could in principle be used to
\textit{develop} a more effective commission-maximizing OTA by treating
the steering RD as a training reward. We mitigate this by (i)~publishing
the governance-parameter phase diagram that attenuates steering alongside
every positive result, (ii)~releasing the symmetric OTA text-audit
protocol and hardened refusal classifier that flag manipulative prose, and
(iii)~framing release language to platform operators, regulators, and
consumer-protection researchers rather than to OTA developers seeking
optimization targets. The net direction of the release is therefore
audit-enabling rather than manipulation-enabling: an OTA developer already
has better signals (real CTR, booking revenue) than our simulator
provides, whereas a regulator does not currently have a standardized audit
instrument.

\textbf{Human subjects.} No human subjects participated in this study. All
traveler-side perception judgments are LLM-generated through guided-JSON
calls to Qwen-14B-AWQ and Llama-3.1-8B. No user data, browsing traces, or
personally identifiable information from real OTA platforms is used,
stored, or released.

\textbf{Synthetic market data.} All bundle prices, commission rates,
hotel/airline inventories, traveler profiles, and vibe archetypes are
generated by the simulator's \texttt{generate\_small\_market} function
with fixed seeds; they do not correspond to, and cannot be reverse-mapped
to, any real OTA's inventory or pricing.

\textbf{Alignment with regulatory instruments.} The benchmark is designed
to produce audit evidence compatible with EU DSA Article~26(3)
(transparency of recommender parameters), FTC 16 CFR Part~255 (endorsement
/ commercial-relationship disclosure), and MiFID~II Article~27
(best-execution-style disclosure). We do not claim regulatory
certification; we provide a reproducible measurement instrument that these
frameworks' compliance-evaluation regimes may adopt or contest.

\textbf{Model licensing and provenance.} Qwen2.5-7B-Instruct and
Qwen2.5-14B-Instruct-AWQ are used under the Tongyi Qianwen license;
Llama-3.1-8B-Instruct is used under the Meta Llama 3.1 Community License.
AWQ weights are consumed as distributed via vLLM \texttt{awq\_marlin}. We
report model versions, commit hashes, and inference configurations in the
Reproducibility Statement below.

\textbf{Pre-registration and audit integrity.} The six-gate thresholds and
the regex v2 hardened refusal classifier were frozen prior to the
Llama-OTA candidate audit (Section~\ref{sec:setup}.7) to avoid
selective-threshold criticism. Source markdown for every audit decision is
released with the code.

\section*{Reproducibility Statement}

\textbf{Release.} TourMart is released at
\url{github.com/usmliuyao/tourmart} and archived at Zenodo
(DOI: \href{https://doi.org/10.5281/zenodo.19709369}{10.5281/zenodo.19709369})
under MIT (code) and CC-BY-4.0 (generated data and audit outputs). The
release contains the market generator with all seeds used in this paper,
the Qwen-7B OTA msgcap pipeline (v4 frozen commit), the paired Phase~1c
replay for both traveler-reader arms, the 36-cell phase-diagram sweep,
the six-gate symmetric audit and regex v2 refusal classifier, paired
raw traveler outputs with extracted perception features, and the four
audit markdown reports referenced in Appendix~\ref{app:sensitivity}.3.

\textbf{Environment and pre-registration.} The OTA producer is
Qwen2.5-7B-Instruct (bf16, vLLM with guided JSON over a per-market
\texttt{bundle\_id} enum); the traveler-reader backbones are
Qwen2.5-14B-Instruct-AWQ and Llama-3.1-8B-Instruct (bf16), both with
guided JSON for the four perception features. All experiments run on a
single RTX~3090 24~GB. Primary sampling seed is~42, with cross-seed
replication at~43 for the OTA-audit. The Round~20 welfare-rule
coefficients, the near-threshold window, the six-gate thresholds, and
the regex v2 refusal classifier were frozen \textit{before} the
Llama-OTA candidate audit was run; the timestamped pre-registration is
included in Appendix~\ref{app:prereg}.

\textbf{One-shot reproduction.} A single shell script in the release
re-runs the $n=143$ paired replay for both arms and regenerates
Table~\ref{tab:1}, Table~\ref{tab:3}, and
Figures~\ref{fig:1}--\ref{fig:3} from cached perceptions in well under
one GPU-hour; a deterministic flag pins vLLM seeds to reproduce reported
numbers within rounding. The statistical pipeline (McNemar exact tests,
95\% RD CIs, Holm-corrected feature-level Wilcoxons, and the
1000-permutation scenario-clustered max-stat procedure) and the
six-gate audit are documented submodules with unit tests, both
rerunnable from the released cached outputs.


\section*{CRediT Author Statement}
\textbf{Yao Liu}: Conceptualization, Methodology, Software,
Validation, Formal analysis, Investigation, Resources, Data Curation,
Writing---Original Draft, Writing---Review \& Editing, Visualization,
Supervision, Project administration, Funding acquisition.

\section*{Declaration of Competing Interests}
The authors declare no competing interests.

\section*{Data Availability}
The TourMart benchmark, all synthetic market scenarios (generated by
\texttt{generate\_small\_market} with fixed seeds), the 818 raw
traveler-LLM call rows per arm (which collapse via the 3-tuple
\texttt{(scenario\_id, traveler\_id, bundle\_id)} key to 143 paired
traveler-LLM outputs with extracted perception features, the four OTA
audit markdown reports, and all statistical analysis code are available
at \url{https://github.com/usmliuyao/tourmart} and archived at Zenodo
(DOI: \href{https://doi.org/10.5281/zenodo.19709369}{10.5281/zenodo.19709369})
under MIT (code) and CC-BY-4.0 (generated data and audit outputs).
A self-contained reproducibility package under
\texttt{reproducibility/} ships pre-computed permutation null
distributions and a \texttt{reproduce\_all.sh} master runner that
regenerates all headline numbers (Table~1, Table~2, Figs.~1--3) from the
paired raw outputs on a standard CPU in under 10~minutes, with no GPU
required. SHA-256 checksums for all input data files are embedded in
\texttt{verify.py}. No proprietary data, human-subject data, or real OTA
inventory data are used in this study.


\appendix

\section{Extended reproducibility notes}
\label{app:repro}

\begin{itemize}
  \item Code: \url{github.com/usmliuyao/tourmart} (Zenodo DOI
        \href{https://doi.org/10.5281/zenodo.19709369}{10.5281/zenodo.19709369}).
  \item Commit hash: [frozen v4].
  \item GPU: 1$\times$ RTX~3090 24~GB (Server~B); Llama-OTA seed-43
        chain also on Server~B.
  \item Additional runtimes: seed-43 msgcap + audit $\approx$ 51~min;
        batch-size confound check (b32 + b128) $\approx$ 8~min.
\end{itemize}

\section{Validity-gate detail}
\label{app:gates}

\subsection*{B.1 Stimulus-audit gates (locked symmetric audit protocol)}

The six audit gates applied symmetrically to each OTA-candidate msgcap
under the regex-based hardened refusal classifier v2:

\begin{table}[t]
\centering
\caption{Symmetric audit gates and rationale.}
\label{tab:b1-gates}
\footnotesize
\setlength{\tabcolsep}{6pt}
\begin{tabular}{@{}lcp{5.2cm}@{}}
\toprule
Gate & Threshold & Rationale \\
\midrule
JSON validity              & $\geq 85\%$  & structured-output survival \\
Bundle\_id coverage        & $\geq 80\%$  & bundle identifier in schema slot \\
Msg word-count median      & $[10, 200]$  & excludes refusals and schema-blob overflow \\
Refusal rate               & $\leq 20\%$  & OTA must produce commission prose \\
Unique-msg ratio (success) & $\geq 30\%$  & template-collapse detection \\
Internal-ID leakage        & $\leq 20\%$  & bid/bundle tokens in user-facing text \\
\bottomrule
\end{tabular}
\end{table}

Gate outcomes by arm:

\begin{table*}[t]
\centering
\caption{Audit gate outcomes per arm (bolded cells are sub-threshold). Verdict
abbreviations: \emph{ref}=refusal, \emph{wc}=word-count, \emph{uq}=uniqueness,
\emph{lk}=ID-leak.}
\label{tab:b1-outcomes}
\scriptsize
\setlength{\tabcolsep}{4pt}
\begin{tabular}{@{}lrrrrrrl@{}}
\toprule
Arm & JSON & Bundle & wc & Refusal & Uniq & ID-leak & Verdict \\
\midrule
Qwen msgcap v4 (s42)        & 100.0\% & 91.3\% & 23   & \textbf{55.9\%} & 74.1\%          & 0.0\%           & FAIL: ref \\
Llama v6 default (s42)      & 100.0\% & 91.7\% & \textbf{6}  & 8.0\%  & \textbf{4.0\%}  & \textbf{84.6\%} & FAIL: wc/uq/lk \\
Llama default (s43)         & 99.3\%  & 91.3\% & \textbf{6}  & 8.5\%  & \textbf{12.5\%} & \textbf{80.9\%} & FAIL: wc/uq/lk \\
Llama probe (s43, verbose)  & 100.0\% & 87.0\% & 21   & 13.0\% & 97.9\%          & 3.7\%           & PASS \\
Llama probe (s42, b=32)     & 100.0\% & 88.9\% & 22.5 & 13.0\% & 97.9\%          & 3.7\%           & PASS \\
Llama probe (s42, b=128)    & 100.0\% & 88.9\% & 23   & 13.0\% & 97.9\%          & 0.0\%           & PASS \\
\bottomrule
\end{tabular}
\end{table*}

Source markdowns:
\path{results/msgcap_v7_v6_vs_qwen_regex_v2.md},
\path{results/msgcap_v7_seed43_default_vs_v6.md},
\path{results/msgcap_v7_seed43_probe_vs_seed43_default.md},
\path{results/msgcap_v7_batch128_vs_batch32.md}.

\subsection*{B.2 Phase~1c calibration diagnostics (per traveler-reader
arm, v4)}

Under the deployed governance point $(\lambda=1, \kappa=0.05)$ and the
diagnostic near-threshold window $[\tau \cdot b - 10\% \cdot b,\ \tau
\cdot b + 5\% \cdot b]$, with 818 traveler calls per arm yielding 143
paired near-threshold episodes (identical scenario sampling across arms):

\begin{table}[t]
\centering
\caption{Phase~1c calibration diagnostics per traveler-reader arm. Parse
and cap-hit are 0/818 missing for both arms; 818 calls per arm.}
\label{tab:b2-calib}
\footnotesize
\setlength{\tabcolsep}{3pt}
\begin{tabular}{@{}lrrrrr@{}}
\toprule
Arm & Parse & Fact acc & Cap-hit & $n$ & orig / fact \\
\midrule
Qwen-14B-AWQ & 100\% & 48.95\% & 0\% & 143 & 56.64\% / 48.95\% \\
Llama-3.1-8B & 100\% & 60.84\% & 0\% & 143 & 64.34\% / 60.84\% \\
\bottomrule
\end{tabular}
\end{table}

No MSG\_ADJ cap-hit is observed at the deployed cap $\kappa=0.05$ under
either reader. Cap-hit rates at larger $\kappa$ multipliers are recovered
from the 2D sensitivity grid (Appendix~\ref{app:sensitivity}.1): hits
emerge only in the saturated regime ($\kappa \in \{20\%, 100\%\}$) where
the transmission effect plateaus.

\subsection*{B.3 Per-feature distributions at the deployed point}

Welfare-rule input features (all 818 calls per arm, both message variants
pooled):

\begin{table}[t]
\centering
\caption{Per-feature distributions (mean, std, saturation
$|v|\geq0.99$).}
\label{tab:b3-features}
\small
\begin{tabular}{lll}
\toprule
Feature & Qwen-14B reader & Llama-3.1-8B reader \\
\midrule
perceived\_fit\_delta & +0.070 (0.242, 3.5\%) & +0.710 (0.147, 0.5\%) \\
perceived\_risk & +0.013 (0.079, 0.5\%) & +0.201 (0.025, 0.0\%) \\
trust\_score & +0.938 (0.229, \textbf{89.9\%}) & +0.857 (0.058, 0.0\%) \\
urgency\_felt & +0.021 (0.105, 0.2\%) & +0.497 (0.087, 0.0\%) \\
\bottomrule
\end{tabular}
\end{table}

\textbf{Saturation warning (Qwen trust\_score, 89.9\% at the upper
ceiling)} is flagged as a construct-validity limitation
(Section~\ref{sec:limitations} item~5 and
Section~\ref{sec:mechanism}.2): the Qwen-reader trust channel is
near-degenerate at the deployed point, which is one reason the dominant
attributable channel in Figure~\ref{fig:2} is \textit{perceived fit}
rather than \textit{trust}. The Llama reader shows no ceiling saturation;
its distributions are tighter and off-center. These cross-reader baseline
shifts---not interventions on any backbone---are the quantitative form of
Section~\ref{sec:mechanism}.2's statement that ``baseline perception
channels differ by traveler-reader backbone.''

\subsection*{B.4 What the gates do NOT guarantee}

The audit gates certify \textit{benchmark eligibility}---that the
OTA-candidate's msgcap is informative enough to drive a transmission
analysis. They do not certify:
\begin{itemize}
  \item ecological realism of message style (deferred to
        marketing-team user study, future work);
  \item absence of steganographic or subtle-persuasion channels the regex
        does not recognize (Section~\ref{sec:limitations} item~6);
  \item that the deployed point $(\lambda=1, \kappa=0.05)$ is optimal for
        any stakeholder; this is the \textit{governance default}, swept by
        the phase diagram in Section~\ref{sec:results}.1.
\end{itemize}

\section{Offline sensitivity analyses}
\label{app:sensitivity}

\subsection*{C.1 2D cap $\times$ coef\_multiplier grid (Fig 1)}
See Figure~\ref{fig:1}.

\subsection*{C.2 Coefficient attribution (Fig 2)}
See Figure~\ref{fig:2}. Holm-corrected exploratory feature-level deltas
are reported in the released supplementary material.

\subsection*{C.3 Llama-OTA cross-seed and batch-size checks}

Four symmetric-stimulus audits under the Llama-3.1-8B-Instruct OTA
candidate:

\textbf{(a) Failure-replication across sampling seeds.} Llama default at
seed~43 vs.\ Llama default at seed~42 (v6): the template-collapse pattern
replicates (both FAIL wc-median, success-only uniqueness, and
internal-ID leakage with near-identical numbers; the top success message
is identical across seeds).

\textbf{(b) Seed-43 style-suffix repair.} Llama default vs.\ Llama probe
(VERBOSE\_PROBE\_SUFFIX) at seed~43: the three failing gates are restored
(wc median $6 \rightarrow 21$, unique-success $12.5\% \rightarrow 97.9\%$,
internal-ID leakage $80.9\% \rightarrow 3.7\%$); the three
originally-passing gates (JSON validity, bundle-id coverage, refusal rate)
remain above threshold (100\%, 87.0\%, 13.0\% respectively).

\textbf{(c) Seed-42 style-suffix repair + batch-size confound check.}
Llama probe at batch=32 vs.\ Llama probe at batch=128, both at seed~42:
all six gates PASS in both batch conditions (wc median 22/23,
unique-success 97.9\%/97.9\%, internal-ID leakage 3.7\%/0.0\%; the three
originally-passing gates also pass), establishing simultaneously that the
suffix repair replicates at seed~42 AND that batch size does not confound
the prompt effect.

\textbf{(d) Producer-side cross-backbone failure-profile comparison.}
Qwen-7B default (Qwen v4) vs.\ Llama-3.1-8B default (Llama v6) under the
regex-based hardened refusal classifier: two distinct failure profiles---Qwen
refusal rate 55.9\% (fails the $\leq 20\%$ refusal gate), Llama
internal-ID leakage 84.6\% (fails the $\leq 20\%$ leakage gate).

Source markdowns:
\path{msgcap_v7_seed43_default_vs_v6.md},
\path{msgcap_v7_seed43_probe_vs_seed43_default.md},
\path{msgcap_v7_batch128_vs_batch32.md},
\path{msgcap_v7_v6_vs_qwen_regex_v2.md}. Full analysis in
\path{refine-logs/PHASE1C_SECTION_5_3_FINDING.md}.

\section{Full pre-registration document (v6 Llama-OTA audit)}
\label{app:prereg}

Reproduced below, unedited except for formatting, from
\path{refine-logs/PHASE1C_V6_LLAMA_OTA_PREREG.md}, frozen 2026-04-20
before the Llama-OTA chain launched on Server~B. \textbf{This is the
pre-registration for the Section~\ref{sec:setup}.7 /
Appendix~\ref{app:sensitivity}.3 Llama-OTA candidate audit only}; the
Round~20 v4 welfare-rule coefficients, near-threshold window, primary
paired-endpoint estimand, and 2D phase-diagram sweep plan
(Section~\ref{sec:formulation}.4, Section~\ref{sec:setup}.3,
Section~\ref{sec:setup}.5) were frozen in a separate earlier
pre-registration at Round~20 and are reproduced in
\path{refine-logs/ROUND_21_FINAL.md} and the locked-config file
\path{configs/round20_v4.yaml}.

\subsection*{D.1 Hypothesis}

The v4 transmission phase diagram (three regimes---attenuated / live /
saturated; peak $+10.49$pp under Qwen-14B reader, $+7.69$pp under
Llama-3.1-8B reader) is \emph{hypothesized} to be a property of
commission-maximizing LLM-OTA prose in general, not an artifact of
Qwen-7B as OTA backbone; the Llama-OTA replay specified in D.2--D.5 is
the pre-registered test of this hypothesis.

\subsection*{D.2 Estimands}

\begin{itemize}
  \item \textbf{Primary (off-diagonal, cross-family)}: Llama-OTA
        $\rightarrow$ Qwen-14B-AWQ reader at deployed $(\lambda=1,
        \kappa=0.05)$: paired risk difference of accept rate between
        commission prose and factual template.
  \item \textbf{Secondary (diagonal)}: Llama-OTA $\rightarrow$
        Llama-3.1-8B reader at the same point, same estimand.
        Pre-registered as lower-weight evidence because of the same-family
        instruction-following prior.
\end{itemize}

\subsection*{D.3 Audit gates (frozen before generation)}

See Appendix~\ref{app:gates}.1 for the six gates. Each gate is identical
to v5 (the deleted Mistral attempt) except for two deliberate loosenings
post-v5: the wc median threshold was widened from $[30, 200]$ to $[10,
200]$ after inspection confirmed 10--30-word single-sentence replies are
informative; the refusal gate was raised from 10\% to 20\% after the Qwen
baseline itself landed at 55.9\% under the hardened regex v2 (documenting
template collapse vs.\ over-hedging as two distinct failure profiles). All
gate changes were applied symmetrically to every arm and re-audited from
raw msgcap output; no arm was tuned retroactively to pass. We are not
aware of a widely-cited dark-pattern or refusal benchmark that fixes a
$20\%$ refusal cutoff, so the widened thresholds are TourMart-internal
documentation choices, not community-standard cutoffs; we discuss the
audit-integrity implications in \S\ref{sec:limitations} (item~13).

\subsection*{D.4 Decision rules (frozen, applied verbatim)}

\begin{itemize}
  \item Off-diagonal RD same sign and within 50\% of v4 $\rightarrow$
        ``cross-family generalization confirmed.''
  \item Off-diagonal null or wrong-sign at both deployed and peak
        $\rightarrow$ ``cross-family dependence; report as
        model-pair-specific.''
  \item Primary passes, secondary fails $\rightarrow$ report $2\times 2$
        mixed; avoid factorial-confirmation language.
  \item Both pass $\rightarrow$ factorial confirmation (not the realized
        outcome; the Llama-reader deployed cell has five discordant pairs
        with an exact McNemar minimum two-sided $p = 0.0625$).
\end{itemize}

\subsection*{D.5 Failure protocol (frozen)}

\begin{itemize}
  \item JSON validity $< 85\%$ $\rightarrow$ chain stops before traveler
        replay. No post-hoc tuning, no fallback to a third OTA model in
        this submission.
  \item Pre-reg-violating configurations (timeout, OOM, sampler drift)
        $\rightarrow$ abort, mark \texttt{.chain\_v6\_failed\_*},
        re-pre-register before re-attempting.
\end{itemize}

\subsection*{D.6 Prior-attempt disclosure}

v5 attempted Mistral-7B-Instruct-v0.3 as the OTA candidate. msgcap
produced $\sim$25\% JSON validity under identical vLLM guided-JSON config
(vs.\ Qwen v4 100\%, Llama v6 100\%). The pre-registered validity gate
(D.3 item~1, $<85\%$ JSON validity $\rightarrow$ abort) killed the chain
\textit{before} traveler replay. No v5 traveler data exists; no v5 figure
was drawn; no v5 number entered the paper. Scripts are retained for
reproducibility (\path{chain_runner_phase1c_v5_mistral.sh},
\path{run_stimulus_audit_v5.py}). Raw artifacts ($\sim$30~MB jsonl +
model weights) were deleted per the frozen failure protocol (D.5 item~2,
``abort and mark \texttt{.chain\_v6\_failed\_*}'') to reclaim disk; the
failure-decision record is preserved in git history and in this appendix.

\subsection*{D.7 Out of scope for this submission}

\begin{itemize}
  \item Additional $(\lambda, \kappa)$ cells beyond the deployed point
        and the v4 sweep grid under the Llama-OTA condition (secondary
        analyses only, not pre-registered).
  \item OTA families beyond Qwen and Llama (no Gemma, no GPT-4o-mini).
  \item Changes to traveler-reader parameters between v4 and v6.
  \item Human-subject evaluation of message ecological validity
        (deferred to marketing-team post-submission user study).
  \item A parallel style-suffix repair for Qwen's over-hedging failure
        profile (flagged as future work;  Claim~2 is stated only for
        Llama's template-collapse profile).
\end{itemize}

\bibliographystyle{elsarticle-num}
\bibliography{references}

@inproceedings{park2023generative,
  title={Generative Agents: Interactive Simulacra of Human Behavior},
  author={Park, Joon Sung and O'Brien, Joseph and Cai, Carrie J. and Morris, Meredith Ringel and Liang, Percy and Bernstein, Michael S.},
  booktitle={Proceedings of the 36th Annual ACM Symposium on User Interface Software and Technology (UIST)},
  year={2023}
}

@inproceedings{aher2023turing,
  title={Using Large Language Models to Simulate Multiple Humans and Replicate Human Subject Studies},
  author={Aher, Gati V. and Arriaga, Rosa I. and Kalai, Adam Tauman},
  booktitle={Proceedings of the 40th International Conference on Machine Learning (ICML)},
  year={2023}
}

@inproceedings{liu2024agentbench,
  title={{AgentBench}: Evaluating {LLMs} as Agents},
  author={Liu, Xiao and Yu, Hao and Zhang, Hanchen and Xu, Yifan and Lei, Xuanyu and Lai, Hanyu and Gu, Yu and Ding, Hangliang and Men, Kaiwen and Yang, Kejuan and others},
  booktitle={International Conference on Learning Representations (ICLR)},
  year={2024}
}

@inproceedings{ma2024agentboard,
  title={{AgentBoard}: An Analytical Evaluation Board of Multi-turn {LLM} Agents},
  author={Ma, Chang and Zhang, Junlei and Zhu, Zhihao and Yang, Cheng and Yang, Yujiu and Jin, Yaohui and Lan, Zhenzhong and Kong, Lingpeng and He, Junxian},
  booktitle={Advances in Neural Information Processing Systems (NeurIPS) Datasets and Benchmarks Track},
  year={2024}
}

@inproceedings{wu2023autogen,
  title={{AutoGen}: Enabling Next-Gen {LLM} Applications via Multi-Agent Conversations},
  author={Wu, Qingyun and Bansal, Gagan and Zhang, Jieyu and Wu, Yiran and Li, Beibin and Zhu, Erkang and Jiang, Li and Zhang, Xiaoyun and Zhang, Shaokun and Liu, Jiale and Awadallah, Ahmed Hassan and White, Ryen W. and Burger, Doug and Wang, Chi},
  booktitle={Conference on Language Modeling (COLM)},
  year={2024},
  eprint={2308.08155},
  archivePrefix={arXiv}
}

@inproceedings{li2023camel,
  title={{CAMEL}: Communicative Agents for ``Mind'' Exploration of Large Language Model Society},
  author={Li, Guohao and Hammoud, Hasan Abed Al Kader and Itani, Hani and Khizbullin, Dmitrii and Ghanem, Bernard},
  booktitle={Advances in Neural Information Processing Systems (NeurIPS)},
  year={2023}
}

@inproceedings{hong2024metagpt,
  title={{MetaGPT}: Meta Programming for a Multi-Agent Collaborative Framework},
  author={Hong, Sirui and Zhuge, Mingchen and Chen, Jiaqi and Zheng, Xiawu and Cheng, Yuheng and Zhang, Ceyao and Wang, Jinlin and Wang, Zili and Yau, Steven Ka Shing and Lin, Zijuan and Zhou, Liyang and Ran, Chenyu and Xiao, Lingfeng and Wu, Chenglin and Schmidhuber, J{\"u}rgen},
  booktitle={International Conference on Learning Representations (ICLR), Oral},
  year={2024},
  eprint={2308.00352},
  archivePrefix={arXiv}
}

@inproceedings{chen2024agentverse,
  title={{AgentVerse}: Facilitating Multi-Agent Collaboration and Exploring Emergent Behaviors},
  author={Chen, Weize and Su, Yusheng and Zuo, Jingwei and Yang, Cheng and Yuan, Chenfei and Chan, Chi-Min and Yu, Heyang and Lu, Yaxi and Hung, Yi-Hsin and Qian, Chen and others},
  booktitle={International Conference on Learning Representations (ICLR)},
  year={2024}
}

@inproceedings{zhou2024sotopia,
  title={{SOTOPIA}: Interactive Evaluation for Social Intelligence in Language Agents},
  author={Zhou, Xuhui and Zhu, Hao and Mathur, Leena and Zhang, Ruohong and Yu, Haofei and Qi, Zhengyang and Morency, Louis-Philippe and Bisk, Yonatan and Fried, Daniel and Neubig, Graham and Sap, Maarten},
  booktitle={International Conference on Learning Representations (ICLR), Spotlight},
  year={2024}
}

@techreport{horton2023homo,
  title={Large Language Models as Simulated Economic Agents: What Can We Learn from {Homo Silicus}?},
  author={Horton, John J. and Filippas, Apostolos and Manning, Benjamin S.},
  institution={National Bureau of Economic Research},
  type={NBER Working Paper},
  number={31122},
  year={2023},
  doi={10.3386/w31122}
}

@inproceedings{li2024econagent,
  title={{EconAgent}: Large Language Model-Empowered Agents for Simulating Macroeconomic Activities},
  author={Li, Nian and Gao, Chen and Li, Mingyu and Li, Yong and Liao, Qingmin},
  booktitle={Proceedings of the 62nd Annual Meeting of the Association for Computational Linguistics (ACL)},
  year={2024}
}

@inproceedings{zhao2024competeai,
  title={{CompeteAI}: Understanding the Competition Dynamics of Large Language Model-based Agents},
  author={Zhao, Qinlin and Wang, Jindong and Zhang, Yixuan and Jin, Yiqiao and Zhu, Kaijie and Chen, Hao and Xie, Xing},
  booktitle={Proceedings of the 41st International Conference on Machine Learning (ICML), Oral},
  year={2024}
}

@article{wang2025recagent,
  title={User Behavior Simulation with Large Language Model-based Agents for Recommender Systems},
  author={Wang, Lei and Zhang, Jingsen and Yang, Hao and Chen, Zhiyuan and Tang, Jiakai and Zhang, Zeyu and Chen, Xu and Lin, Yankai and Song, Ruihua and Zhao, Wayne Xin and Xu, Jun and Dou, Zhicheng and Wang, Jun and Wen, Ji-Rong},
  journal={ACM Transactions on Information Systems (TOIS)},
  year={2025}
}

@inproceedings{piatti2024govsim,
  title={Cooperate or Collapse: Emergence of Sustainable Cooperation in a Society of {LLM} Agents},
  author={Piatti, Giorgio and Jin, Zhijing and Kleiman-Weiner, Max and Sch{\"o}lkopf, Bernhard and Sachan, Mrinmaya and Mihalcea, Rada},
  booktitle={Advances in Neural Information Processing Systems (NeurIPS)},
  year={2024}
}

@inproceedings{abdelnabi2024stakeholders,
  title={Cooperation, Competition, and Maliciousness: {LLM}-Stakeholders Interactive Negotiation},
  author={Abdelnabi, Sahar and Gomaa, Amr and Sivaprasad, Sarath and Sch{\"o}nherr, Lea and Fritz, Mario},
  booktitle={Advances in Neural Information Processing Systems (NeurIPS) Datasets and Benchmarks Track},
  year={2024}
}

@article{mathur2019darkpatterns,
  title={Dark Patterns at Scale: Findings from a Crawl of 11K Shopping Websites},
  author={Mathur, Arunesh and Acar, Gunes and Friedman, Michael J. and Lucherini, Elena and Mayer, Jonathan and Chetty, Marshini and Narayanan, Arvind},
  journal={Proceedings of the ACM on Human-Computer Interaction},
  volume={3},
  number={CSCW},
  articleno={81},
  year={2019},
  doi={10.1145/3359183}
}

@inproceedings{gray2018darkpatterns,
  title={The Dark (Patterns) Side of {UX} Design},
  author={Gray, Colin M. and Kou, Yubo and Battles, Bryan and Hoggatt, Joseph and Toombs, Austin L.},
  booktitle={Proceedings of the 2018 CHI Conference on Human Factors in Computing Systems (CHI)},
  year={2018},
  doi={10.1145/3173574.3174108}
}

@inproceedings{digeronimo2020uidark,
  title={{UI} Dark Patterns and Where to Find Them: A Study on Mobile Applications and User Perception},
  author={Di Geronimo, Linda and Braz, Larissa and Fregnan, Enrico and Palomba, Fabio and Bacchelli, Alberto},
  booktitle={Proceedings of the 2020 CHI Conference on Human Factors in Computing Systems (CHI)},
  year={2020},
  doi={10.1145/3313831.3376600}
}

@inproceedings{gray2024ontology,
  title={An Ontology of Dark Patterns Knowledge: Foundations, Definitions, and a Pathway for Shared Knowledge-Building},
  author={Gray, Colin M. and Santos, Cristiana Teixeira and Bielova, Nataliia and Mildner, Thomas},
  booktitle={Proceedings of the 2024 CHI Conference on Human Factors in Computing Systems (CHI)},
  year={2024},
  doi={10.1145/3613904.3642436}
}

@article{luguri2021shining,
  title={Shining a Light on Dark Patterns},
  author={Luguri, Jamie and Strahilevitz, Lior Jacob},
  journal={Journal of Legal Analysis},
  volume={13}, number={1}, pages={43--109},
  year={2021},
  doi={10.1093/jla/laaa006}
}

@inproceedings{patro2020fairrec,
  title={{FairRec}: Two-Sided Fairness for Personalized Recommendations in Two-Sided Platforms},
  author={Patro, Gourab K. and Biswas, Arpita and Ganguly, Niloy and Gummadi, Krishna P. and Chakraborty, Abhijnan},
  booktitle={Proceedings of The Web Conference 2020 (WWW)},
  year={2020},
  doi={10.1145/3366423.3380196}
}

@misc{ftc2022darkpatterns,
  title={Bringing Dark Patterns to Light: {FTC} Staff Report},
  author={{Federal Trade Commission}},
  year={2022}, month={September},
  howpublished={\url{https://www.ftc.gov/reports/bringing-dark-patterns-light}},
  note={P214800}
}

@misc{ftc2023endorsement,
  title={Guides Concerning the Use of Endorsements and Testimonials in Advertising, 16 {CFR} Part 255},
  author={{Federal Trade Commission}},
  year={2023}, month={July},
  howpublished={\url{https://www.ecfr.gov/current/title-16/chapter-I/subchapter-B/part-255}},
  note={Final rule effective July 26, 2023; 88 FR 48092}
}

@misc{eu2022dsa,
  title={Regulation ({EU}) 2022/2065 on a Single Market for Digital Services (Digital Services Act)},
  author={{European Parliament and Council}},
  year={2022}, month={October},
  howpublished={Official Journal of the European Union L 277/1}
}

@misc{eu2014mifid2,
  title={Directive 2014/65/{EU} (MiFID II), Article 27: Obligation to Execute Orders on Terms Most Favourable to the Client},
  author={{European Parliament and Council}},
  year={2014},
  howpublished={Official Journal of the European Union L 173/349}
}

@article{reviglio2024recommender,
  title={The Regulation of Recommender Systems Under the {DSA}: A Transition from Default to Multiple and Dynamic Controls?},
  author={Reviglio, Urbano and Fabbri, Matteo},
  journal={DSA Observatory Policy Analysis},
  year={2024},
  month={November},
  howpublished={\url{https://dsa-observatory.eu/2024/11/22/the-regulation-of-recommender-systems-under-the-dsa-a-transition-from-default-to-multiple-and-dynamic-controls/}}
}

@book{benshahar2014disclosure,
  title={More Than You Wanted to Know: The Failure of Mandated Disclosure},
  author={Ben-Shahar, Omri and Schneider, Carl E.},
  publisher={Princeton University Press},
  year={2014},
  isbn={9780691161709}
}

@article{loewenstein2014disclosure,
  title={Disclosure: Psychology Changes Everything},
  author={Loewenstein, George and Sunstein, Cass R. and Golman, Russell},
  journal={Annual Review of Economics},
  volume={6},
  pages={391--419},
  year={2014},
  doi={10.1146/annurev-economics-080213-041341}
}

@article{aridor2022recommenderoriginals,
  title={Recommenders' Originals: The Welfare Effects of the Dual Role of Platforms as Producers and Recommender Systems},
  author={Aridor, Guy and Gon{\c{c}}alves, Duarte},
  journal={International Journal of Industrial Organization},
  volume={83},
  pages={102845},
  year={2022},
  doi={10.1016/j.ijindorg.2022.102845}
}

@inproceedings{xie2024travelplanner,
  title={{TravelPlanner}: A Benchmark for Real-World Planning with Language Agents},
  author={Xie, Jian and Zhang, Kai and Chen, Jiangjie and Zhu, Tinghui and Lou, Renze and Tian, Yuandong and Xiao, Yanghua and Su, Yu},
  booktitle={Proceedings of the 41st International Conference on Machine Learning (ICML), Spotlight},
  year={2024},
  eprint={2402.01622}
}

@article{chen2024travelagent,
  title={{TravelAgent}: An {AI} Assistant for Personalized Travel Planning},
  author={Chen, Aili and Ge, Xuyang and Fu, Ziquan and Xiao, Yanghua and Chen, Jiangjie},
  journal={arXiv preprint arXiv:2409.08069},
  year={2024}
}

@article{chaudhry2025tripcraft,
  title={{TripCraft}: A Benchmark for Spatio-Temporally Fine Grained Travel Planning},
  author={Chaudhuri, Soumyabrata and Purkar, Pranav and Raghav, Ritwik and Mallick, Shubhojit and Gupta, Manish and Jana, Abhik and Ghosh, Shreya},
  journal={arXiv preprint arXiv:2502.20508},
  year={2025},
  eprint={2502.20508},
  archivePrefix={arXiv},
  primaryClass={cs.CL}
}

@inproceedings{yang2025widehorizon,
  title={Wide-Horizon Thinking and Simulation-Based Evaluation for Real-World {LLM} Planning with Multifaceted Constraints},
  author={Yang, Dongjie and others},
  booktitle={Advances in Neural Information Processing Systems (NeurIPS)},
  year={2025},
  eprint={2506.12421}
}

@article{qian2025tripscore,
  title={{TripScore}: Benchmarking and Rewarding Real-World Travel Planning with Fine-Grained Evaluation},
  author={Qu, Yincen and Xiao, Huan and Li, Feng and Li, Gregory and Zhou, Hui and Dai, Xiangying and Dai, Xiaoru},
  journal={arXiv preprint arXiv:2510.09011},
  year={2025},
  eprint={2510.09011},
  archivePrefix={arXiv},
  primaryClass={cs.CL}
}

@article{cheng2025travelbench,
  title={Beyond Itinerary Planning---A Real-World Benchmark for Multi-Turn and Tool-Using Travel Tasks},
  author={Cheng, Xiang and Hu, Yulan and Zhang, Xiangwen and Xu, Lu and Tan, Lide and Pan, Zheng and Li, Xin and Liu, Yong},
  journal={arXiv preprint arXiv:2512.22673},
  year={2025},
  eprint={2512.22673},
  archivePrefix={arXiv},
  primaryClass={cs.AI}
}

@inproceedings{duetting2024mechanism,
  title={Mechanism Design for Large Language Models},
  author={Duetting, Paul and Mirrokni, Vahab and Paes Leme, Renato and Xu, Haifeng and Zuo, Song},
  booktitle={Proceedings of the ACM Web Conference 2024 (WWW), Best Paper},
  year={2024},
  doi={10.1145/3589334.3645511}
}

@inproceedings{soumalias2024rag,
  title={Truthful Aggregation of {LLMs} with an Application to Online Advertising},
  author={Soumalias, Ermis and Curry, Michael J. and Seuken, Sven},
  booktitle={International Conference on Learning Representations (ICLR)},
  year={2025},
  eprint={2405.05905},
  archivePrefix={arXiv},
  primaryClass={cs.GT}
}

@inproceedings{dubey2024summaries,
  title={Auctions with {LLM} Summaries},
  author={Dubey, Kumar Avinava and Feng, Zhe and Kidambi, Rahul and Mehta, Aranyak and Wang, Di},
  booktitle={Proceedings of the 30th ACM SIGKDD Conference on Knowledge Discovery and Data Mining (KDD)},
  year={2024},
  pages={713--722},
  publisher={ACM},
  doi={10.1145/3637528.3672022}
}

@article{fish2024collusion,
  title={Algorithmic Collusion by Large Language Models},
  author={Fish, Sara and Gonczarowski, Yannai A. and Shorrer, Ran I.},
  journal={arXiv preprint arXiv:2404.00806},
  year={2024}
}

@inproceedings{sharma2024sycophancy,
  title={Towards Understanding Sycophancy in Language Models},
  author={Sharma, Mrinank and Tong, Meg and Korbak, Tomasz and Duvenaud, David and Askell, Amanda and Bowman, Samuel R. and Cheng, Newton and Durmus, Esin and Hatfield-Dodds, Zac and Johnston, Scott R. and Kravec, Shauna and Maxwell, Timothy and McCandlish, Sam and Ndousse, Kamal and Rausch, Oliver and Schiefer, Nicholas and Yan, Da and Zhang, Miranda and Perez, Ethan},
  booktitle={International Conference on Learning Representations (ICLR)},
  year={2024},
  eprint={2310.13548}
}

@misc{durmus2024persuasion,
  title={Measuring the Persuasiveness of Language Models},
  author={Durmus, Esin and Lovitt, Liane and Tamkin, Alex and Ritchie, Stuart and Clark, Jack and Ganguli, Deep},
  year={2024}, month={April},
  publisher={Anthropic},
  howpublished={\url{https://www.anthropic.com/news/measuring-model-persuasiveness}}
}

@article{salvi2025persuasiveness,
  title={On the Conversational Persuasiveness of {GPT-4}},
  author={Salvi, Francesco and Horta Ribeiro, Manoel and Gallotti, Riccardo and West, Robert},
  journal={Nature Human Behaviour},
  volume={9}, pages={1645--1653},
  year={2025},
  doi={10.1038/s41562-025-02194-6}
}

@article{salvi2026commercial,
  title={Commercial Persuasion in {AI}-Mediated Conversations},
  author={Salvi, Francesco and Cuevas, Alejandro and Horta Ribeiro, Manoel},
  journal={arXiv preprint arXiv:2604.04263},
  year={2026},
  eprint={2604.04263},
  archivePrefix={arXiv},
  primaryClass={cs.CY}
}

@article{bansal2025magentic,
  title={Magentic Marketplace: An Open-Source Environment for Studying Agentic Markets},
  author={Bansal, Gagan and Hua, Wenyue and Huang, Zezhou and Fourney, Adam and Swearngin, Amanda and Epperson, Will and Payne, Tyler and Hofman, Jake M. and Lucier, Brendan and Singh, Chinmay and Mobius, Markus and Nambi, Akshay and Yadav, Archana and Gao, Kevin and Rothschild, David M. and Slivkins, Aleksandrs and Goldstein, Daniel G. and Mozannar, Hussein and Immorlica, Nicole and Murad, Maya and Vogel, Matthew and Kambhampati, Subbarao and Horvitz, Eric and Amershi, Saleema},
  journal={arXiv preprint arXiv:2510.25779},
  year={2025},
  eprint={2510.25779},
  archivePrefix={arXiv},
  primaryClass={cs.AI}
}

@article{hackenburg2024microtargeting,
  title={Evaluating the Persuasive Influence of Political Microtargeting with Large Language Models},
  author={Hackenburg, Kobi and Margetts, Helen},
  journal={Proceedings of the National Academy of Sciences (PNAS)},
  volume={121}, number={24}, pages={e2403116121},
  year={2024},
  doi={10.1073/pnas.2403116121}
}

@article{bai2023aipersuasion,
  title={{LLM}-generated messages can persuade humans on policy issues},
  author={Bai, Hui and Voelkel, Jan G. and Muldowney, Shane and Eichstaedt, Johannes C. and Willer, Robb},
  journal={Nature Communications},
  volume={16},
  pages={6037},
  year={2025},
  doi={10.1038/s41467-025-61345-5}
}

@article{matz2024personalized,
  title={The Potential of Generative {AI} for Personalized Persuasion at Scale},
  author={Matz, Sandra C. and Teeny, Jacob D. and Vaid, Sumer S. and Peters, Heinrich and Harari, Gabriella M. and Cerf, Moran},
  journal={Scientific Reports},
  volume={14}, pages={4692},
  year={2024},
  doi={10.1038/s41598-024-53755-0}
}

@article{rogiers2024persuasion,
  title={Persuasion with Large Language Models: A Survey},
  author={Rogiers, Alexander and Noels, Sander and Buyl, Maarten and De Bie, Tijl},
  journal={arXiv preprint arXiv:2411.06837},
  year={2024}
}

@inproceedings{kran2025darkbench,
  title={{DarkBench}: Benchmarking Dark Patterns in Large Language Models},
  author={Kran, Esben and Nguyen, Hieu Minh and Kundu, Akash and Jawhar, Sami and Park, Jinsuk and Jurewicz, Mateusz Maria},
  booktitle={International Conference on Learning Representations (ICLR), Oral},
  year={2025},
  eprint={2503.10728},
  archivePrefix={arXiv}
}

@inproceedings{fanous2025syceval,
  title={{SycEval}: Evaluating {LLM} Sycophancy},
  author={Fanous, Aaron and Goldberg, Jacob and Agarwal, Ank and Lin, Joanna and Zhou, Anson and Xu, Sonnet and Bikia, Vasiliki and Daneshjou, Roxana and Koyejo, Sanmi},
  booktitle={Proceedings of the AAAI/ACM Conference on AI, Ethics, and Society (AIES)},
  year={2025},
  volume={8},
  number={1},
  pages={893--900},
  doi={10.1609/aies.v8i1.36598},
  eprint={2502.08177},
  archivePrefix={arXiv}
}

@inproceedings{ersoy2026darkwebagents,
  title={Investigating the Impact of Dark Patterns on {LLM}-Based Web Agents},
  author={Ersoy, Devin and Lee, Brandon and Shreekumar, Ananth and Arunasalam, Arjun and Ibrahim, Muhammad and Bianchi, Antonio and Celik, Z. Berkay},
  booktitle={Proceedings of the 47th IEEE Symposium on Security and Privacy (S\&P)},
  year={2026},
  eprint={2510.18113},
  archivePrefix={arXiv},
  primaryClass={cs.CR},
  note={To appear}
}

@article{argyle2023outofone,
  author  = {Argyle, Lisa P. and Busby, Ethan C. and Fulda, Nancy and Gubler, Joshua R. and Rytting, Christopher and Wingate, David},
  title   = {Out of One, Many: {U}sing Language Models to Simulate Human Samples},
  journal = {Political Analysis},
  volume  = {31},
  number  = {3},
  pages   = {337--351},
  year    = {2023},
  publisher = {Cambridge University Press},
  doi     = {10.1017/pan.2023.2}
}

@inproceedings{sandvig2014auditing,
  author    = {Sandvig, Christian and Hamilton, Kevin and Karahalios, Karrie and Langbort, Cedric},
  title     = {Auditing Algorithms: {R}esearch Methods for Detecting Discrimination on Internet Platforms},
  booktitle = {64th Annual Meeting of the International Communication Association ({ICA}), Data and Discrimination Preconference},
  year      = {2014}
}

@inproceedings{hannak2013measuring,
  author    = {Hannak, Aniko and Sapiezynski, Piotr and Molavi Kakhki, Arash and Krishnamurthy, Balachander and Lazer, David and Mislove, Alan and Wilson, Christo},
  title     = {Measuring Personalization of Web Search},
  booktitle = {Proceedings of the 22nd International Conference on World Wide Web ({WWW})},
  pages     = {527--538},
  year      = {2013},
  publisher = {ACM},
  doi       = {10.1145/2488388.2488435}
}

@article{metaxa2021auditing,
  author  = {Metaxa, Dana{\"e} and Park, Joon Sung and Robertson, Ronald E. and Karahalios, Karrie and Wilson, Christo and Hancock, Jeffrey and Sandvig, Christian},
  title   = {Auditing Algorithms: {U}nderstanding Algorithmic Systems from the Outside In},
  journal = {Foundations and Trends in Human-Computer Interaction},
  volume  = {14},
  number  = {4},
  pages   = {272--344},
  year    = {2021},
  publisher = {now Publishers},
  doi     = {10.1561/1100000083}
}

@article{kim2021ota,
  author  = {Kim, Woo Gon and Pillai, Sweta and Haldorai, Karthik and Ahmad, Wajeeha},
  title   = {Dark Patterns Used by Online Travel Agency Websites},
  journal = {Annals of Tourism Research},
  volume  = {88},
  pages   = {103055},
  year    = {2021},
  publisher = {Elsevier},
  doi     = {10.1016/j.annals.2020.103055}
}

@inproceedings{wu2022jmefairness,
  author    = {Wu, Haolun and Mitra, Bhaskar and Ma, Chen and Diaz, Fernando and Liu, Xue},
  title     = {Joint Multisided Exposure Fairness for Recommendation},
  booktitle = {Proceedings of the 45th International ACM SIGIR Conference on Research and Development in Information Retrieval ({SIGIR})},
  pages     = {703--714},
  year      = {2022},
  publisher = {ACM},
  doi       = {10.1145/3477495.3532007}
}

@misc{china2022algorec,
  title  = {Provisions on the Administration of Algorithmic Recommendations of Internet Information Services ({CAC} Order No.~9)},
  author = {{Cyberspace Administration of China and Ministry of Industry and Information Technology and Ministry of Public Security and State Administration for Market Regulation}},
  year   = {2022}, month = {March},
  howpublished = {\url{http://www.cac.gov.cn/2022-01/04/c_1642894606364259.htm}},
  note   = {Effective 2022-03-01; in force when this paper was prepared}
}

@misc{eu2019p2b,
  title  = {Regulation ({EU})~2019/1150 of the European Parliament and of the Council of 20 June 2019 on Promoting Fairness and Transparency for Business Users of Online Intermediation Services},
  author = {{European Parliament and Council}},
  year   = {2019}, month = {July},
  howpublished = {Official Journal of the European Union L 186/57}
}

\end{document}